\documentclass[namedreferences]{solarphysics}
\usepackage[optionalrh]{spr-sola-addons} 
\usepackage{graphicx}        
\usepackage[usenames,dvipsnames,svgnames]{xcolor}
\usepackage{hyperref}

\newcommand{\m}[1]{#1}
\newcommand{\mm}[1]{#1}

\newcommand{\BE}{\begin{equation}}
\newcommand{\EE}{\end{equation}}
\newcommand{\BA}{\begin{eqnarray}}
\newcommand{\EA}{\end{eqnarray}}
 \newcommand{\fig}[1]{Figure~\ref{fig:#1}}

 \newcommand{\sect}[1]{Section~\ref{sect:#1}}

\newcommand{\Jz}{J_z}

\newcommand{\eg}{\textit{e.g.}}

\newcommand{\ie}{\textit{i.e.}}
\newcommand{\insitu}{{\it in-situ}}

\newcommand{\Halpha}{H$\alpha$}

\begin{document}

\begin{article}

\begin{opening}

\title{From coronal observations to MHD simulations, the building blocks for 3D models of solar flares}

\author{M.~\surname{Janvier}$^{1}$\sep
        G.~\surname{Aulanier}$^{2}$\sep
        P.~\surname{D\'emoulin}$^{2}$
               }
\runningauthor{M. Janvier et al.}
\runningtitle{Characteristics of solar flares}

   \institute{
 $^{1}$ Division of Mathematics, University of Dundee, Dundee DD1 4HN, Scotland, United Kingdom\\
        email: \url{mjanvier@maths.dundee.ac.uk} \\ 
 $^{2}$ Observatoire de Paris, LESIA, UMR 8109 (CNRS), F-92195 Meudon Principal Cedex, France \\
              }

\begin{abstract}
Solar flares are energetic events taking place in the Sun's atmosphere, and \m{their} effects can greatly impact the environment of the surrounding planets. In particular, eruptive flares, \m{as opposed to} confined flares, launch coronal mass ejections \m{into} the interplanetary medium, and \m{as such, are} one of the main drivers of space weather.
After briefly reviewing the main characteristics of solar flares, we summarize the processes that can \m{account} for the \m{build up and release of energy} during their evolution.
In particular, we focus on the development of recent 3D numerical simulations \m{that} explain many of the observed flare features. These simulations can also \m{provide} predictions of the dynamical evolution of coronal and photospheric magnetic \m{field.  Here we present a few observational examples that, together with numerical modelling, point to the underlying physical mechanisms of the eruptions.}

\end{abstract}

\end{opening}

\section{Introduction}
     \label{sect:Introduction} 

Solar flares are sudden brightenings occurring in the atmosphere of our Sun that can be monitored in a large range of the electromagnetic spectrum. They are associated with intense radiations, the release of energetic particles, as well as in some cases the ejection of solar plasma and magnetic field \m{in} the form of coronal mass ejections (hereafter, CMEs).
As such, solar flares are major drivers of space weather \citep{Gosling1991} as they can impact planetary environments \citep{Prange2004} and in particular human activities \citep{Schrijver2014,Lugaz2015}. Further away from our solar system, flares are also seen to occur on other stars \citep{Linsky2000}, as reported by the new findings from the NASA-Kepler mission \citep{Walkowicz2011,Wichmann2014}.
Therefore, understanding the mechanisms of solar flares is important, not only to better predict the changes occurring in the Earth's environment, but also to a large extent, to understand star-planet interactions. This will \m{ultimately} help to assess planetary habitability conditions \citep{Poppenhaeger2014, Strugarek2014}.

The first solar flare ever recorded was also one of the most powerful (although its energy content has been \m{greatly} debated, see \citealt{McCracken2001,Tsurutani2003,Wolff2012,Cliver2013}). \m{It} occurred on 1 September 1859, as reported by Carrington and Hodgson \citep{Carrington1859,Hodgson1859}. To date, the \m{flare of} 4 November 2003 \m{is} the most energetic flare recorded with recent instrumentation. \m{Its soft X-ray emission was} above the X28 range. The existence of super-flares on other stars \citep{Maehara2012} therefore requires investigating whether flares more intense than that observed so far can occur on our aging Sun \citep{Schrijver2012,Aulanier2013}. This question, linked with the dynamo evolution in the Sun's interior, also requires a further understanding on how the energy of flares is stored, and what mechanisms can \m{account} for its release. Note that flares are seen to occur \m{over} a large range of the energy spectrum: small-scale flares, associated with rather low energy \m{events} (10$^{16}$ to 10$^{20}$J), are monitered in the quiet regions of the Sun. These so-called nanoflares or subflares will not be the focus of the present article.

With ever increasing spatial and temporal resolutions, instruments onboard more recent spacecraft now monitor the Sun in different wavelengths, providing \m{detailed observations of the evolution} of the photospheric, chromospheric and coronal plasma and magnetic field. They therefore provide strong constraints on models of solar flares. 
As such, understanding their mechanisms encompasses a variety of approaches, from deciphering the observational features to developing numerical models that capture their underlying physics. Dedicated reviews of solar flares are
numerous: while some review the relevant physics to describe the magnetic conversion process during flares (see \eg\ \citealt{Priest2002,Shibata2011}), others report on the variety of their observational features (see \eg\ \citealt{Krucker2008,Fletcher2011}). Although reviews on the models and simulations developed to explain flares are also available \citep[\eg\ ][]{Chen2011}, they generally focus on a 2D or 2.5D description of the magnetic field in the solar atmosphere. To fill this gap, we \m{present} in the following \m{sections} a review of the recent modelling developments that incorporate the intrinsic, and therefore essentially 3D nature of solar flares. We will see the wealth of methods now developed to extrapolate the coronal magnetic field, to characterise 3D \mm{topological} features, and to reproduce the dynamics of the field in 3D. The present review aims at explaining
 how these techniques and observations provide the building blocks to construct 3D models of solar flares. In particular, we propose \mm{in the final section how different features of eruptive flares can be combined to construct a comprehensive, yet rather complete}, standard 3D model for eruptive flares.

\sect{Observations} \m{briefly reviews} the observational characteristics of solar flares, with a particular focus on the differences between confined and eruptive flares. \sect{Energy} reviews the current understanding of energy storage and \m{introduces} the concept of topology to understand where flare energy release \mm{is initiated}. \sect{flareevolution} \m{discusses} the evolution of flares, from their \m{onset through their decline}, and in \sect{eruptiveflares} we focus on eruptive flares leading to the ejection of flux ropes and the characteristics of the 3D magnetic reconnection process. Finally, conclusions to this article are presented in  \sect{Conclusion}.

\section{Observational Characteristics of Confined and Eruptive Flares} 
      \label{sect:Observations}      

\subsection{Flare Emissions} 
      \label{sect:emission}      

  \begin{figure}  
 \centerline{ \includegraphics[width=1\textwidth]{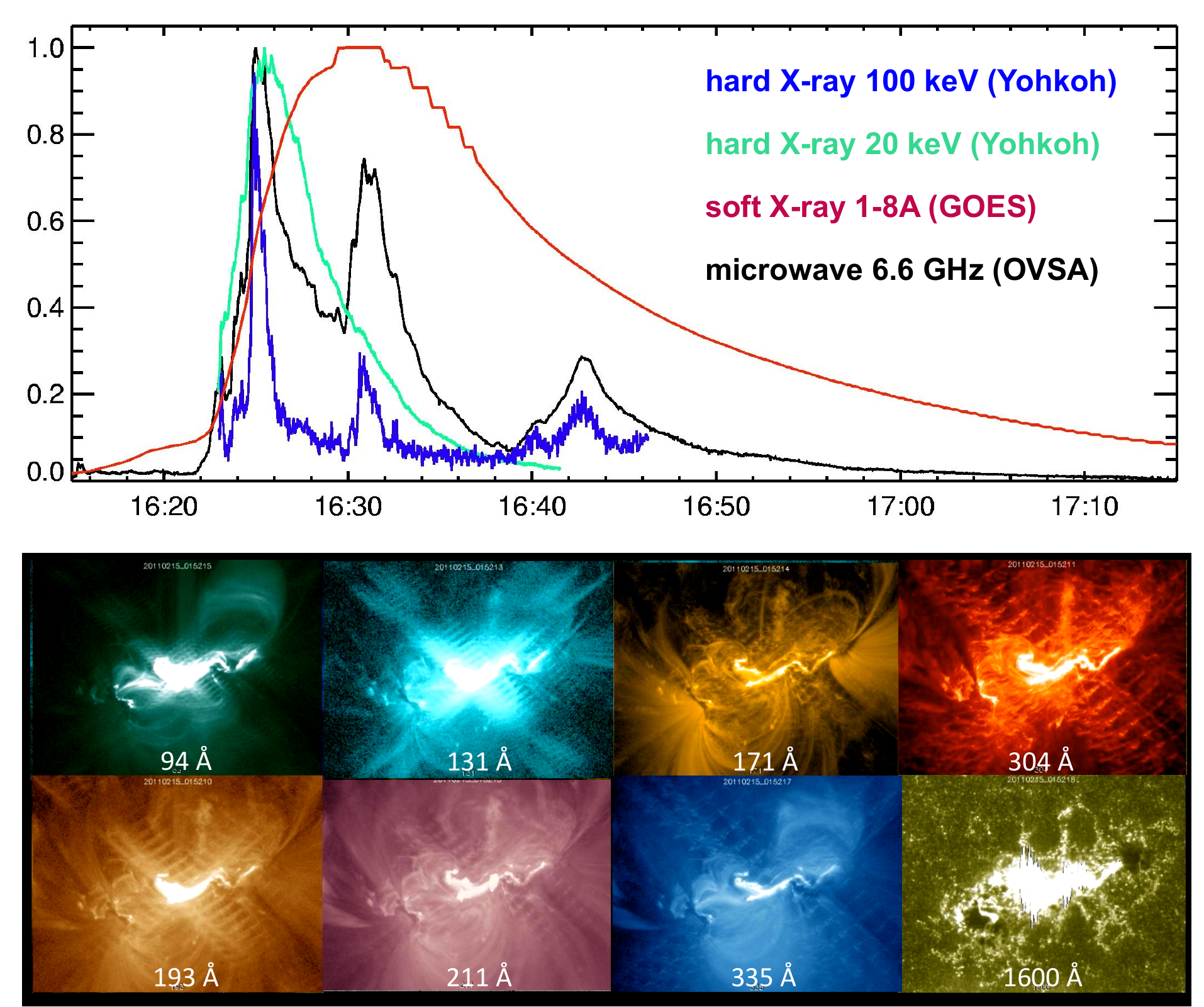} }
\caption{\textbf{Top}: Emissions of the 19 October 2001 X1.6 flare (\mm{SOL2001-10-19T16:30})
observed with different instruments. The hard X-ray observations were made with the HXT/Yohkoh \citep{Kosugi1991}, the soft X-ray observations with the GOES spacecraft, and finally, the microwave observations come from the Owens Valley Solar Array (OVSA, \citealt{Gary1990}). The figure is adapted from \citet{Qiu2009}. \textbf{Bottom}: the 15 February 2011 X2.2 flare (\mm{SOL2011-02-15T01:56}) in AR 11198 (at 01:45 UT) as observed in the different filters of the AIA/SDO instrument. The seven EUV channels (from 94~\AA\ to 335~\AA) are separated into the ionised iron-dominated coronal channels (first 3 columns) and the He II-dominated chromosphere 304~\AA, while the 1600~\AA\ in the EUV shows the emission in the upper photosphere. The \m{ribbons are} most clearly visible at 335~\AA\ \m{while the flare loops are seen at a variety of wavelengths (\eg\ 94~\AA\ and 131~\AA)} emitting at different temperatures (adapted from \citealt{Schrijver2011}).}
 \label{fig:emission}
\end{figure}  

Flares are defined as impulsive releases of \m{radiative energy over} a large range of the electromagnetic spectrum  (see \fig{emission}). The classification of flares is made with the soft X-ray, \m{using the} 1-8~\AA\ band of the GOES-5 satellite (\textit{Geostationary Orbiting Environmental Satellites}).
This classification ranks flares from the A class to the X class (A,B,C,M,X), with the X class associated with the most energetic flares (with a GOES flux exceeding $10^{-4}$ \mm{Wm}$^{-2}$ at Earth, and the other classes decreasing by one order of magnitude).
Weak events occurring at an energy level about $10^{-9}$ times lower than large flares have been \mm{named} ``nanoflares'', and are being investigated as a possible mechanism for the heating of the corona \citep[see][and references therein]{Klimchuk2014}, following the seminal idea of \citet{Parker1988}.
In between, microflares, with energies of order $10^{-6}$ times those of large flares, have been statistically investigated by \inlinecite{Hannah2011}. The authors, using the \textit{Reuven Ramaty High Energy Solar Spectroscopic Imager} (RHESSI, \citealt{Lin2002}), showed that the physics that underly larger flare events can be extended down to smaller events. 
This is a remarkable outcome of the survey of flares and their emission within a large range of the electromagnetic spectrum. Indeed, the distribution of flare parameters, such as their estimated radiated energy, can be represented by a power law (see \eg\ Figure 3 of \citealt{Schrijver2012} and references in the paper).

  \begin{figure}  
 \centerline{ \includegraphics[width=1\textwidth]{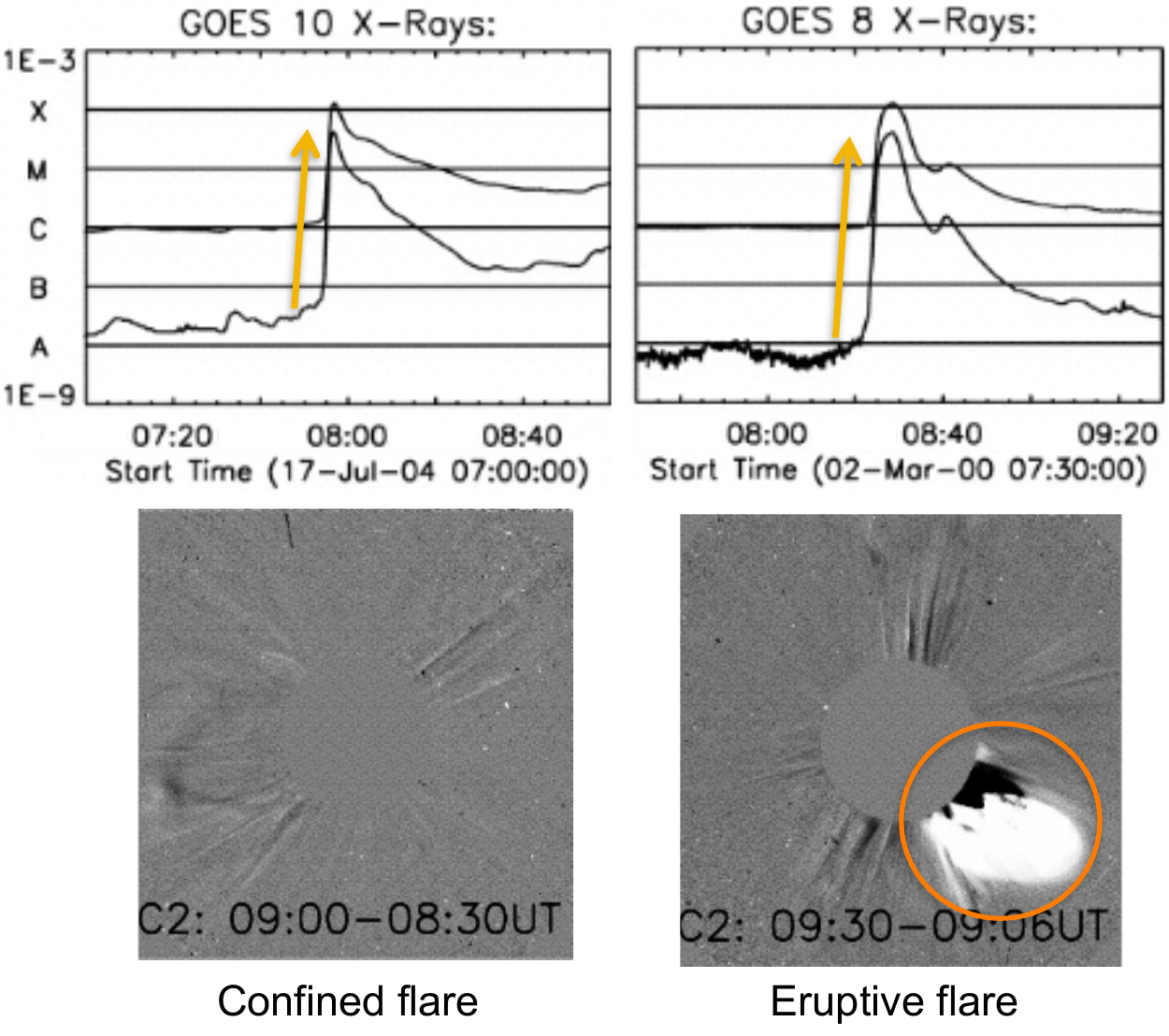} }
\caption{An example of two flares showing a similar GOES Soft X-ray evolution (top) with a similar impulsive and decay phase time scale, while the running difference coronagraph images (bottom) show the large-scale effect of an eruptive flare with the launch of a CME (adapted from \citealt{Wang2007}).
}
 \label{fig:confinederuptive}
\end{figure}  

A distinction can be made between confined and eruptive flaring events, a classification that was proposed already back in the 1980s by \inlinecite{Svestka1986}. Confined flares are generally impulsive in time and compact in space, and the flare loops (see \sect{flareloops}) \m{contain} most of the energy. \m{They are} accompanied \m{by} a population of accelerated particles and intense plasma radiation indicating that strong heating occurs. Eruptive flares, \m{on} the contrary, generally extend to a large volume of the corona and lead to the ejection of solar material \m{in} the form of coronal mass ejections. Although a distinction cannot be made between those two flare categories by only looking at the light curves, a coronagraph is generally a straightforward way to define them, as shown in \fig{confinederuptive}.
Note that if eruptive flares are defined by the presence of a CME, the inverse is not true: CMEs have been seen leaving the solar atmosphere without an associated flare emission. These cases, referred to as stealth CMEs \citep[see the review of][and references therein]{Howard2013}, can be associated with coronal phenomena invisible to present instruments, and therefore should be considered as a space weather prediction pitfall, as they are less well monitored.

\inlinecite{Yashiro2005} investigated the link between CMEs and flares by associating events observed over a period of 7 years, and found that the CME association rate with the intensity of the flare increases from 20\% for C-class flares (between C3 and C9 levels) to 100\% for large flares (above the X3 level). However, X-class flares are not necessarily all associated with CMEs as the recent series of X-class flares of October 2014 in AR 12192 has shown (\citet{Thalmann2015}, see also the studies by \citealt{Feynman1994} and \citealt{Green2002}).  This therefore leaves the question of the relation between intense flares and eruptiveness still open.

Both eruptive and confined events share some similarities in the evolution of the flare emission, as they are both characterized by an increase in radiation and are associated with emissions due to a population of non-thermal electrons, that is generally not detected in the non-flaring corona. However, eruptive flares are generally associated with long duration events, while confined flares typically have a much shorter time scale, as was shown in the study of \citet{Yashiro2006}.
The flare emissions for both type of events in different wavelengths do not peak at the same time \citep[see also][and references therein]{Fletcher2011}: typically, a brightening in soft X-rays of a few keV is first detected, associated with the chromospheric plasma  heated to coronal temperatures. Also, radio emissions in the metric range, such as type III radio bursts caused by escaping electron beams, give an indication that the pre-flare phase is not entirely thermal.
Most of the flare energy is then released during a short-span phase referred to as the impulsive phase, that is highly non-thermal. It is generally associated with hard X-ray emissions due to  energetic electrons, as well as gamma-ray line emission caused by energetic ions. It is believed that most of the free coronal energy is released during this phase under the form of non-thermal particles, with electrons reaching energies up to 300 MeV, and ions up to 10 GeV.
 As the flare progresses during this phase, the soft X-ray emission increases in size and total flux, while the lower chromosphere also radiates more, as \m{indicated} by a growing \Halpha\ emission.
After the impulsive phase the decay phase starts, where most of the emissions due to non-thermal particles have disappeared. The hot plasma decreases in temperature to the point that it also completely disappears, and flare loops, filled with evaporated dense plasma, emit in \Halpha.

\subsection{Flare loops, Ribbons and Kernels} 
      \label{sect:flareloops}      

\begin{figure}  
 \centerline{ \includegraphics[width=1\textwidth]{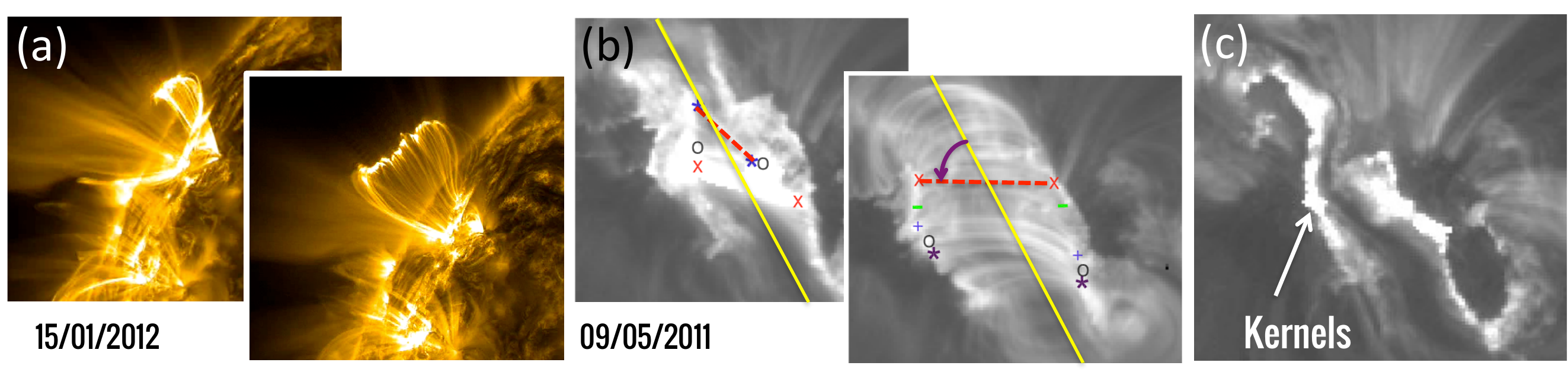} }
\caption{An example of the typical evolution of flare loops. 
  (a) Low-to-high-altitude evolution of flare loops as seen at 171~\AA\ in the 15/01/2012 flare (\mm{SOL2012-01-15T23:59})  with the AIA/SDO \citep{Lemen2012}. 
  (b) Strong-to-weak shear evolution of flare loops as seen at 195~\AA\ in the 11/05/2011 flare observed with the EUVI/STEREO A \citep{Wuelser2004}. The shear angle is defined between the photospheric inversion line (PIL, yellow line) of the photospheric magnetic field and the line joining the two footpoints of one loop (red line).
  (c) The two ribbons appearing in the early phase of the 09/05/2011 (\mm{SOL2011-05-09T20:59})  flare show clumps of brightenings, or kernels (adapted from \citealt{Aulanier2012}). }
 \label{fig:loopsribbons}
\end{figure}

One of the most remarkable features of flares is the appearance of very hot coronal loops (T$\ge$10 MK), referred to as flare loops, which remain above the flaring region (contrary to the CME during eruptive flares). Since they emit in different temperature ranges, it is possible to see their evolution in time with different filters. Looking at this evolution in just one filter is however misleading, as these loops would be seen growing in size (\fig{loopsribbons}a). However, following the same loop, it typically appears first in hot filters and progressively shifts to colder filters while it shrinks in height \citep{Forbes1996}.
Moreover, newly formed loops are forming above already existing loops \citep[\eg\ ][]{Raftery2009, Sun2013}.

Another characteristics of these flare loops is the evolution of their shear (\fig{loopsribbons}b). The shear angle is  
defined between the polarity inversion line (PIL) and the line that joins their footpoints, and as such, requires flares with a well-defined two-ribbon configuration and that are mostly observed near center-disk (so as to avoid any projection effects). This evolution was studied in various papers \citep{Asai2003,Su2006,Aulanier2012}, where conjugate footpoints at first indicated strongly sheared field lines, and later reveal less sheared coronal connections (\fig{loopsribbons}b).  \citet{Su2007} also studied 50 X- and M-class flares observed with TRACE, 
associated with an arcade of flare loops, and found that 86\% of these flares showed a general decrease in the shear angle. 

The flare ribbons, discussed above, are intense emissions forming at the footpoints of flare loops, and are generally well seen before the appearance of the flare loops, mostly in \Halpha, but also in other EUV filters, as shown in \fig{loopsribbons}b,c. \mm{Flare ribbons appear in their fully developed state as coherent and elongated structures. They are frequently accompanied by spatially localised and strongly brightened clumps}, referred to as kernels (\fig{loopsribbons}c).
Because of their strong emission in chromospheric lines, it is believed that kernels and ribbons, \mm{form as a result of strong heating of the chromospheric plasma, most likely by collisions with non-thermal electrons}. The evaporation of this strongly heated plasma can be seen as Doppler shifts in the diagnostics of spectral lines, for example with the EIS/HINODE instrument, as investigated by \citet{delZanna2006}, \citet{Milligan2009} and \citet{Young2013}. \mm{Some parts of} flare ribbons are also associated with the sites of HXR emissions (see \citealt{Fletcher2001b}, \citealt{Asai2002}, \citealt{Reid2012}). 

Eruptive flares generally display two ribbons with a large portion generally parallel to each other, and which also present a hook shape at their ends (\fig{loopsribbons}c, see also \citealt{Chandra2009}). It was shown, \eg\ in \citet{Janvier2014}, that this hook surrounds a dimming region generally associated with the footpoints of the ejected CME. For confined flares, flare ribbons can be more numerous \citep[see for example][]{Wang2014} or have different shapes \citep[see for example a circular ribbon configuration in][]{Masson2009}. As such, \mm{the presence of two ribbons with a $J$-shape structure is interpreted as an evidence of an eruptive flare}.
    
\subsection{Filaments/Prominences, and CMEs} 
      \label{sect:CMEs}      

  \begin{figure}  
 \centerline{ \includegraphics[width=1\textwidth]{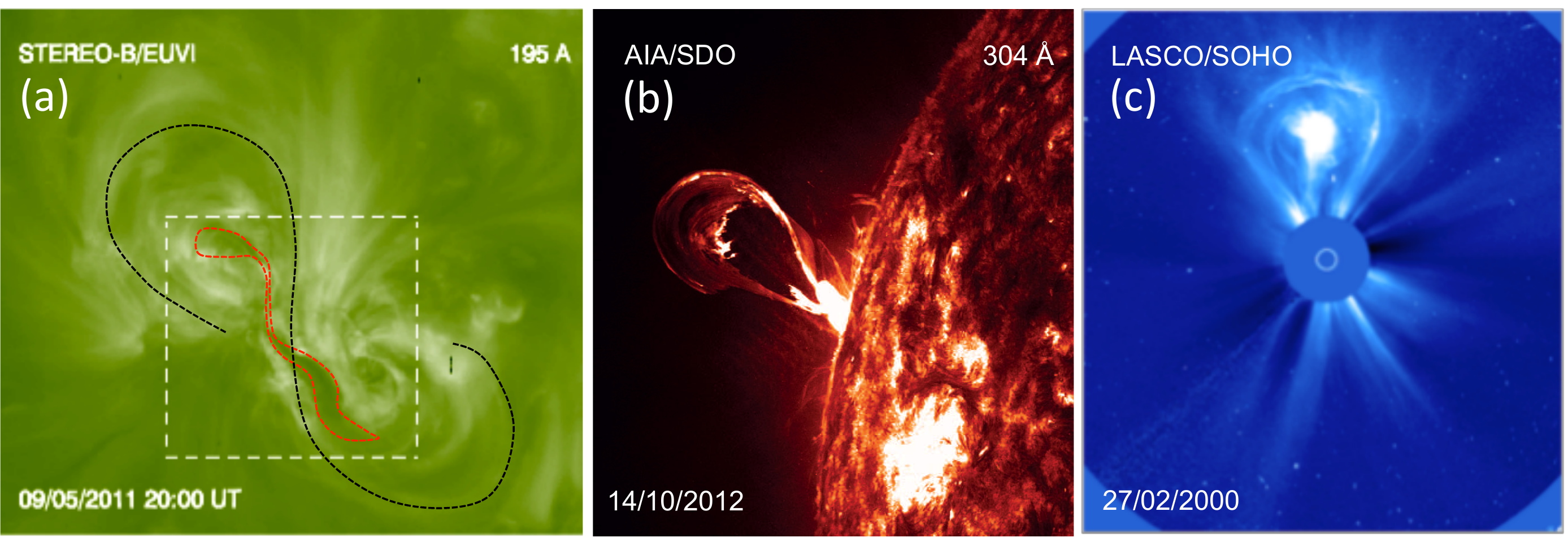} }
\caption{(a) Flare precursors can be seen prior to eruptions in some active regions. Here, the AR 11193, which led to a C-class flare on 09/05/2011, is shown one hour before the event. It features a filament (indicated with the red dashed lines) and a sigmoidal structure (black dashed lines). 
  (b) The early phase of a CME eruption during the 14 October 2012 flare, accompanying an intense emission lower in the atmosphere, as seen with AIA/SDO. 
  (c) A CME seen by the coronagraph LASCO onboard SOHO.
}
 \label{fig:filamentCME}
\end{figure}  

Prominences, which appear on the limb as bright structures, or filaments, dark filamentary objects on disk, are potential progenitors of eruptive flares (\fig{filamentCME}a). These structures, which are observationally complex to understand (see the recent reviews of \citealt{Schmieder2012, Engvold2015} and accompanying references), are indeed often associated with CMEs (\fig{filamentCME}b,c). This association has been thoroughly investigated in a series of papers (\citealt{Subramanian2001}, see also the reviews of \citealt{Schmieder2014} and \citealt{Webb2015}).
In particular, filaments/prominences \mm{eruptions and CMEs are generally well associated, from 56\% to 92\% of the former associated with CMEs} \citep{Hori2002, Gopalswamy2003, Jing2004}.

Note that some flares, which would be confined by definition (no CME observed), clearly show the beginning of a prominence or filament eruption, although instead of lifting off to \mm{greater} altitude, the structure remains above the region. Such cases are generally \mm{called} failed eruptions (see for example in \citealt{Ji2003}, \citealt{Gilbert2007} and \citealt{Joshi2015}). 

In active regions, CME precursors are often seen as a bright structures made of $S$ or $J$-shaped coronal loops, forming what is called a sigmoid \citep{Rust1996,Canfield1999,Green2009}. They are present with the filament (see the example in \fig{filamentCME}a), and they are generally well seen in the X-ray range (see the catalogue of sigmoids in \citealt{Savcheva2014}).
 
\section{Energy Build-up and Release} 
      \label{sect:Energy}      

The energy of a solar flare is typically estimated \m{to be} between $10^{28}$ to a few $10^{32}$erg \citep[see][]{Schrijver2012}. 
With a plasma $\beta <1$, the magnetic field drives the coronal activity, and its associated energy is the only component that can \m{account for flare power}. Looking at the evolution of light curves (for example those of \fig{confinederuptive}), several questions \m{arise}: How to explain the long duration energy storage phase, lasting from a few days to a few weeks? What are the mechanisms for the sudden energy release? What triggers flares, and how is the energy converted into other types of energy (\ie, kinetic energy, radiation)? In the following, we review recent theoretical and numerical developments that try to answer these questions.

\subsection{Mechanisms for Energy Build-up} 
      \label{sect:Buildup}      

For a given distribution of the normal component of a photospheric field, the potential field (current free) has the lowest magnetic energy.
\m{Any} excess is called ``free energy" and it is stored as currents in non-potential field \m{configurations.  Indeed, there is indirect observational evidence of}
strong electric currents stored in magnetic structures prior to an eruption onset. This is \m{particularly} the case with the X-ray emitting sigmoids,  
as investigated by \eg\ \citet{Green2007, McKenzie2008, Savcheva2012b}, and with the presence of twisted filaments/prominences \citep[\eg\ ][]{Williams2005, Koleva2012}.
How can such magnetic \m{fields that store their energy in aligned} currents be created?

\begin{figure}  
 \centerline{ \includegraphics[width=1\textwidth]{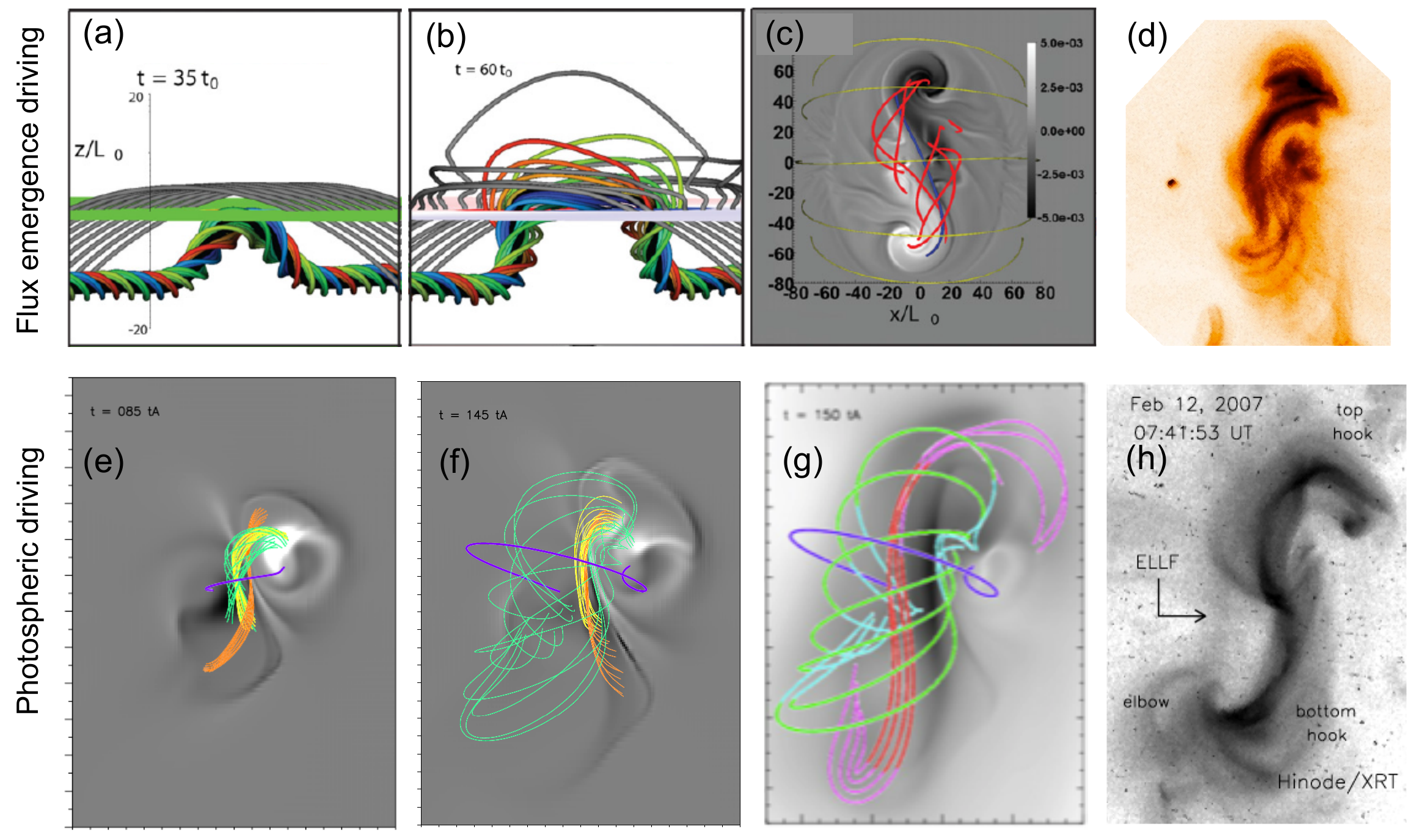} }
\caption{Two examples of 3D MHD simulations forming twisted flux tubes in the corona. (a,b) \citet{Leake2013} proposed a flux rope emergence simulation. The twisting motion of the polarities resulting from the emergence leads to current-carrying coronal loops. These loops are represented with the red lines in panel (c), while the background represents the normalized vertical current density $J_z$.  The shapes of the red lines, especially the elongated one that runs almost parallel to the inversion line, are quite similar to the $J$/$S$-shape coronal loops found in the sigmoid of 12 February 2007 \citep[panel (d)][]{McKenzie2008}. (e,f) The simulation of \citet{Aulanier2010} forms a flux rope structure via photospheric motion (twist) and diffusion at the photosphere (panels e and f). This creates a set of field lines running above the inversion line (in green and orange) and a sigmoidal distribution of currents (background grey levels). The sigmoidal field lines and the integrated current (panel g) also reproduces well the shape of the 12/02/2007 sigmoid {(panel h).}
} 
 \label{fig:currents}
\end{figure}

Two mechanisms can \m{account} for the energy storage. One is the emergence of sub-photospheric current-carrying flux tubes from the convection zone (\fig{currents}, top row). Several authors have reported observational evidence that the Sun's magnetic field is often twisted on emergence {\citep{Leka1996, vanDriel1997, Luoni2011}.} 

Numerical simulations are heavily used to study the photospheric emergence since the physical problem is complex (see the reviews by \opencite{Fan2009,Hood2012,Cheung2014}). This complexity is in particular linked to the presence of a large stratification of the plasma and magnetic field (since the scale height \mm{of the photosphere} is only about 150 km).  This implies that a flux rope coming from the convective zone is no longer buoyant below the photosphere. There, its magnetic field is blocked in a flat structure until enough magnetic flux has accumulated. This leads to the development of the so-called Parker instability, allowing the magnetic flux to emerge by pieces into the chromosphere and corona. \mm{As such, the magnetic field is completely changed from its initial state during this emergence}, and finally it is a new flux rope that reforms in the corona \citep{Hood2009,Toriumi2012}.  In some simulations, the coronal flux rope is eventually ejected \citep[\eg\ ][]{Archontis2012, Kusano2012}. However, the key parameters leading to such an ejection still remain to be understood (see \sect{flareevolution}).

Another mechanism that can \m{account} for the formation of a current-carrying magnetic field is slow photospheric motions, for example by twisting the polarities, or by inducing a shearing motion parallel to the inversion line.  Moreover, flux ropes can form by reconnection of low field lines, such as in the model of \citet{vanBallegooijen1989}. This reconnection at the PIL occurs as the magnetic field diffuses gradually at the photosphere as found in numerical simulations \citep[\eg\ ][]{Amari2003b,Aulanier2010,Amari2011, Amari2014}.
There, the presence of current density layers at bald patches (regions where the magnetic field is tangent to the photosphere) leads to the reconnection of magnetic field lines that create a newly formed structure made of twisted flux bundles, embedded in an overlying, almost potential arcade. High electric currents confined in this structure are similar to the $J$- or $S$-shaped sigmoids often observed in observations (see \fig{currents}, bottom row).

\subsection{Reconnection: a Mechanism for Flares} 
      \label{sect:reconnection}      

  \begin{figure}  
 \centerline{ \includegraphics[width=1\textwidth]{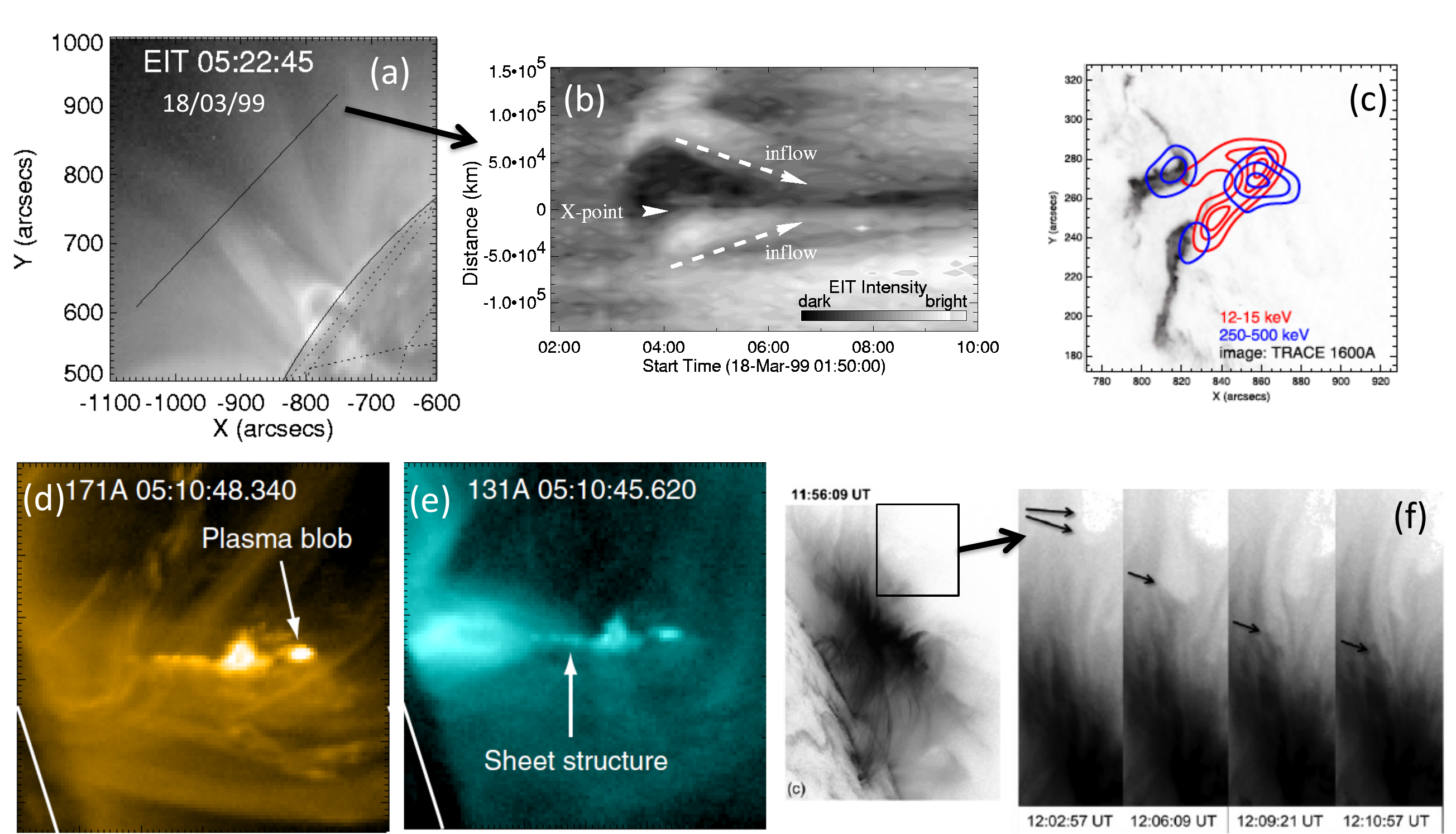} }
\caption{Examples of reconnection signatures in the solar corona.
(a,b) Converging motions (inflows) toward a thin sheet-like region, above flare loops during the 18 March 1999 (\mm{SOL1999-03-18T04:04}) event as observed by EIT/SOHO \citep{Yokoyama2001}. Panel (b) is the time evolution of the emission in 195~\AA\ along the slit shown in (a).
(c) HXR emissions from RHESSI during the 20 January 2005 event \citep[\mm{SOL2005-01-20T06:30,}][]{Krucker2008}, showing the presence of a strong emission source ($>$250 keV, blue) above the top of the loop (red contours).  (d,e) Observation of a plasmoid ejection during the 18 August 2010 flare (\mm{SOL2010-08-18T05:48}) and its associated reconnection region, seen within two filters of AIA/SDO instrument \citep{Takasao2012}. (f) Observation of a supra-arcade downflow seen as a void propagating sunward in the region above the flare loops, during the 22 October 2011 event  (\mm{SOL2011-10-22T10:00}) in the AIA 131~\AA\ filter \citep{Savage2012}
} 
 \label{fig:recoevidences}
\end{figure}  

\m{As discussed in the previous section,} magnetic reconnection is thought to be the key mechanism \mm{that permits the evolution of the field} \m{in the storage phase and the eruptive phase of flares}. 
\m{Already recognizing its importance} back in the 1950s, \inlinecite{Sweet1958b} and \inlinecite{Parker1957} developed a theory, based on a work by \inlinecite{Dungey1953}, that could explain how the magnetic energy contained in the corona could be released efficiently while giving the intense radiation recorded during flares. 
Such a mechanism is related to the fact that in an evolving magnetic configuration \m{with a null point}, a current sheet typically \m{forms. This thin sheet introduces small scales 
where} dissipative effects become important.
In such a current sheet, the magnetic energy is dissipated into heat, kinetic energy and energetic particles.
By applying this theory to solar flares, they defined a flaring configuration with four domains of connectivity. The reconnection of field lines \m{between these regions} converts magnetic energy into other forms of energy, while transforming the magnetic field configuration.

Since these pioneering works, there has been a tremendous number of studies on magnetic reconnection, \m{especially} in relation to flares, and the reader is referred to previous reviews \citep[][Chap. 12]{Forbes2010, Shibata2011,Priest2014}. Here, we only note that in 2D, \ie\ in the presence of separatrices, which are special field lines separating different domains of connectivity, the definition of reconnection is clear. It can be defined both mathematically by a change of connectivity domain for some field lines, and physically by the dissipation of the magnetic energy within thin layers. A definition that is both mathematical and physical is an important distinction to make, as in 3D, mathematically well-defined separations of connectivity domains do not necessarily exist (see \sect{QSLs}), while the reconnection, defined physically as the dissipation of magnetic energy in thin electric current layers, remains a valid concept.

\m{There is much} evidence that the reconnection process takes place in the corona. One aspect of it is the formation of a population of non-thermal particles, that is also observed in plasma experiments and in \insitu\ data of the magnetosphere (see chapter V.B in the review of \citealt{Yamada2010}). For solar flares, several authors have reported on the existence of hard-X ray sources above the flare loop tops \citep[\eg\ ][]{Masuda1994, Hudson2001, Sui2003}, indicative of a source forming high-energy particles (see \fig{recoevidences}c). In particular, \citet{Masuda1994} proposed a heating process due to high speed jets produced by reconnection, colliding with the top of the reconnected flare loops. Furthermore, flare loops are ordered in temperature with the outermost loops being the hottest, as expected \m{for} a reconnection process \m{that forms} them sequentially at larger height \citep[see \eg][]{Forbes1996,Tsuneta1996}.

Another strong evidence that magnetic reconnection is taking place is given \m{by} the presence of inflows and outflows from \m{and to} the reconnection \m{site}. For example, as reported by \citet{Yokoyama2001}, the authors used EUV and soft X-ray observations of a flare with the EIT/SOHO \citep{Delaboudiniere1995} and SXT/Yohkoh \citep{Tsuneta1991} instruments, and pointed out the pulling process of coronal loops in a region underneath a plasmoid ejection and above the top of the flare loops (\fig{recoevidences}a,b). Using diagnostics from the soft-X ray emissions, they were able to deduce an inflow speed. Outflows, on the other hand, are often seen as void structures in EUV and SXR emission ranges above flare loops. Sunward-flowing voids are referred to as supra-arcade downflows (SADs, see \citet{McKenzie2011} and \fig{recoevidences}f), propagating with a speed ranging from 50 to 500 km.s$^{-1}$ (sub-Alfv\'enic). They have been interpreted as coronal loop cross section evacuated from the reconnection region \citep{Savage2012}. Inversely, plasmoids ejected away from the Sun have also been reported \citep{Nishizuka2010} in increasing numbers with the temporal and spatial capacity improvements of AIA onboard SDO (\citet{Takasao2012} and \fig{recoevidences}d,e). Using radio emissions, more refined analysis of the dynamics of these plasmoids \m{provides} more insights on the reconnection mechanism at the origin of their ejection \citep[\eg\ ][]{Nishizuka2015}. Finally, inflows and outflows are often seen together, as reported in the coronal observations of \citet{Yokoyama2001,Li2009,Savage2012b,Takasao2012,Su2013}.

When magnetic energy is released in the corona, a significant part of this energy is also transported along field lines by thermal conduction fronts, and/or energetic particles, \mm{and/or Poynting flux (\textit{e.g.} \citealt{Fletcher2008}, \citealt{Birn2009}, \citealt{Longcope2012})} toward the lower atmosphere. This generally makes the first energy release of flares \citep[see][and references therein]{Krucker2005,SaintHilaire2005}.
Then, the chromospheric plasma becomes hot (up to tens of millions of degrees, as seen in soft X-rays), as non-thermal particles collide with the particles of this atmospheric layer. Intense radiation can be seen as hard X-ray sources at the footpoints of \mm{bright} flare loops, \mm{which used to be referred to as post-flare loops}, (\eg\ kernels, see \sect{flareloops}), but also at the footpoints of twisted \mm{loops}, as reported by \citet{Liu2009} and \citet{Guo2012}. Note that for both cases, HXR sources were found at the endpoints of the associated filaments, appearing during a failed eruption. Eruptive flares do not usually show such intense emissions, although it is possible to find such cases \citep{Musset2015}. As the chromosphere expands, evaporation of hot plasma ensues, filling up coronal loops that are then seen as flare loops (see \sect{flareloops}).
Much of the energy coming from the non-thermal population is eventually converted into thermal energy and radiation, seen \mm{dominantly as the UV and optical component of the flare \citep[see \eg][]{Milligan2014}}. 
 Therefore, measuring the radiative losses of flares  is almost equivalent to the flare's energy release. Almost, since one would also need to measure the bulk kinetic energy from the ejecta and the escaping particle populations to obtain an accurate description of the total energy of a flare.

\subsection{Using Topology to Find Reconnection Regions} 
      \label{sect:Toporeco}      

In order to localize where reconnection could occur, one needs to search for the regions where the evolving magnetic field \m{can} form current layers, \m{in other words,} where ideal MHD breaks down. These regions are found by analyzing the magnetic topology of the field.

The first models of reconnection were 2D. In 2D, the magnetic field can be separated in different domains of connectivity by special field lines that are called separatrices. 
Two separatrices cross at a null point, where the magnetic field vanishes \citep{Sweet1958a}.
From the 1990s, a great deal of work has been made with extensions of those mathematically defined features from 2D to 3D.
Separatrices in 3D are surfaces, which delimit different domains of connectivity.  The intersection of two separatrices defines a separator, which is a privileged location for reconnection to occur. It is a special field line joining two null points.
The complete topology description of a magnetic field is given by the skeleton formed by null points, separators and separatrices. Reconnection taking place at each of these topological features in 3D has been at the core of several works which explored the complexity of possible magnetic topologies \citep[\eg\ ][and references therein]{Longcope1996,Parnell1996,Priest1996,Pontin2004,Pontin2005,Pontin2007,Priest2009,Parnell2010}.

The application of the above theory to solar flares started to be developed around the 1980s by placing a few magnetic sources below the photosphere, so as to reproduce the observed magnetograms. \m{The topology is then computed and compared} to flare observations \citep{Baum1980,Gorbachev1988,Lau1993}. 
The complexity of the model was later increased to include many more sources \m{with} positions and strengths \m{being}  computed by a least square fit of the model to the magnetogram of a specific observation.
This procedure defines the coronal magnetic field and its topology (via magnetic connectivity between magnetic sources). Several flare studies found that the observed flare ribbons are located in the vicinity of the computed separatrices. This is evidence that the energy release occurs by magnetic reconnection (\eg\ \opencite{Mandrini1993}; \opencite{Demoulin1994}).

This approach was further developed in the so-called Magnetic Charge Topology (MCT) model 
\citep[\eg\ ][]{Barnes05,Longcope2005}, in particular by estimating the amount of electric current flowing along the separator before reconnection and the amount of reconnected flux during the observed flare. 
Although successful in describing reconnection signatures for a variety of solar active regions, the MCT model is \m{limited} since
it requires \m{one} to first split the observed magnetogram \m{into} a series of separated polarities, which is ambiguous for unipolar regions. \m{The} topology is computed as if every polarity, modeled by a charge, is surrounded by  
flux free regions. The above approach introduces an ensemble of null points and separatrices \m{that} are not all in the coronal field computed from the photospheric magnetogram \m{using} force-free extrapolation methods.

\subsection{Quasi-Separatrix Layers} 
      \label{sect:QSLs}      

Flares with comparable characteristics are observed with and without magnetic null points and their associated separatrices \citep{Demoulin1994b}.
Therefore, extending the concept of separatrices to regions where the magnetic field connectivity remains continuous but changes with a strong gradient,\inlinecite{Demoulin1996} introduced the concept of quasi-separatrix layers (QSLs). These are mathematically defined with the Jacobi matrix of connectivity  transformation from a photospheric positive polarity domain to a negative polarity domain, or simply put, by the field line mapping of the magnetic field. Regions where the magnetic field is strongly distorted, or ``squashed'', are defined by a strong value of the squashing degree $Q$ \citep{Titov2002,Pariat2012}.  
As such, computing QSLs allows the magnetic topology of an active region \mm{to be defined} when the coronal configuration is computed by any means (\eg\ magnetic extrapolation or MHD simulation).

QSLs are a generalisation in 3D of the concept of separatrices. In the limit where photospheric unipolar regions are separated by flux-free regions, or when a magnetic null enters in the coronal domain, or when a pair of nulls appears after a bifurcation, the central parts of QSLs become separatrices \citep{Demoulin1996,Restante2009}. QSLs with large $Q$ values still remain on both side of a separatrix \citep[where $Q$ becomes infinite,][]{Masson2009}.
Finally, because of the strong distortion of the magnetic field, strong current density build-up typically occurs at QSLs during the magnetic field evolution (as shown in \citealt{Aulanier2005,Buchner2006,Effenberger2011,Janvier2013}). As such, QSLs are preferential locations for reconnection to take place. This was evidenced observationally by the correspondence between flare ribbons and QSL locations and by the presence of concentrated electric currents located at the border of the QSLs \citep{Demoulin1997, Mandrini1997}.   

\citet{Hesse1988} and \citet{Schindler1988} developed a general framework for 3D reconnection in the absence of null point and separatrices.  Their formulation of the magnetic field with Euler potentials allows the description of the evolution of field lines and to pinpoint how and where they are cut, therefore accessing the physics of 3D reconnection.  It occurs where localised non-ideal regions are set. QSLs are the locations where these non-idealness can occur as strong current density typically build there during an evolution. Therefore, the two approaches are complementary, with the QSLs providing the large scale topology, localising where reconnection occurs, and the formalism of Hesse and Schindler describing the local reconnection process within the MHD framework \citep{Demoulin1996b,Richardson2012}.   These results were recently extended to kinetic numerical simulations \citep{Wendel2013,Finn2014}.
 
The QSL theory has been tested in several simple configurations, for example in cases where two interacting bipoles in a solar active region could be easily identified \citep{Demoulin1997}. 
The theory has also been extensively used to understand more complex cases. 
A sample of the many other studies looking at the location of reconnection signatures, typically flare ribbons, and comparing them successfully with the locations of QSLs, can be found in \citet{Schmieder1997,Bagala2000,Wang2000,Masson2009,Chandra2011,Savcheva2012,Zhao2014,Dudik2014,Dalmasse2015}.
This close correspondence between flare ribbons and separatrix/QSLs found in various magnetic configurations, and derived from the photospheric magnetograms, is strong evidence that magnetic reconnection is releasing the free magnetic energy in flares.

In the above flare applications, computing QSLs in a magnetic field volume is more demanding \m{of} computer ressources than locating the null points in the MCT model, due to the large amount of field lines to \m{process.  The advantage of the QSL method} is that it can be applied to any magnetic extrapolation technique, and with any theoretical model (for example see \sect{QSLsfluxropes}). Furthermore, \m{the calculation of} QSLs can be sped up with the use of a refinement algorithm using an adaptive grid that progressively computes field lines only where the thin QSLs are located \citep{Demoulin1996,Pariat2012}.

\section{Magnetic Environment of Flares} 
      \label{sect:flareevolution}      

\subsection{Confined Flares: Emerging Flux} 
      \label{sect:emergingflux}      

A model for emerging flux and energy dissipation was proposed by \cite{Heyvaerts1977} then \cite{Forbes1984} for confined flares: as new flux emerges from the photosphere, it presses against overlying field structures, leading to the formation of a current density layer. As this process goes on, the current in the sheet grows, until a critical threshold is reached and triggers a micro-instability. The occurence of such an instability can lead to an increase in the local resistivity of the plasma, which in turn can rapidly dissipate the current (see \sect{reconnection}).
Then, not only is flux emergence associated with the transfer of current-carrying magnetic field structures (see \sect{Buildup}) from the solar convection zone to the atmosphere, it can also account for triggering energy releases by forcing the transformation of the magnetic field configuration. 

 \begin{figure}  
 \centerline{ \includegraphics[width=1\textwidth]{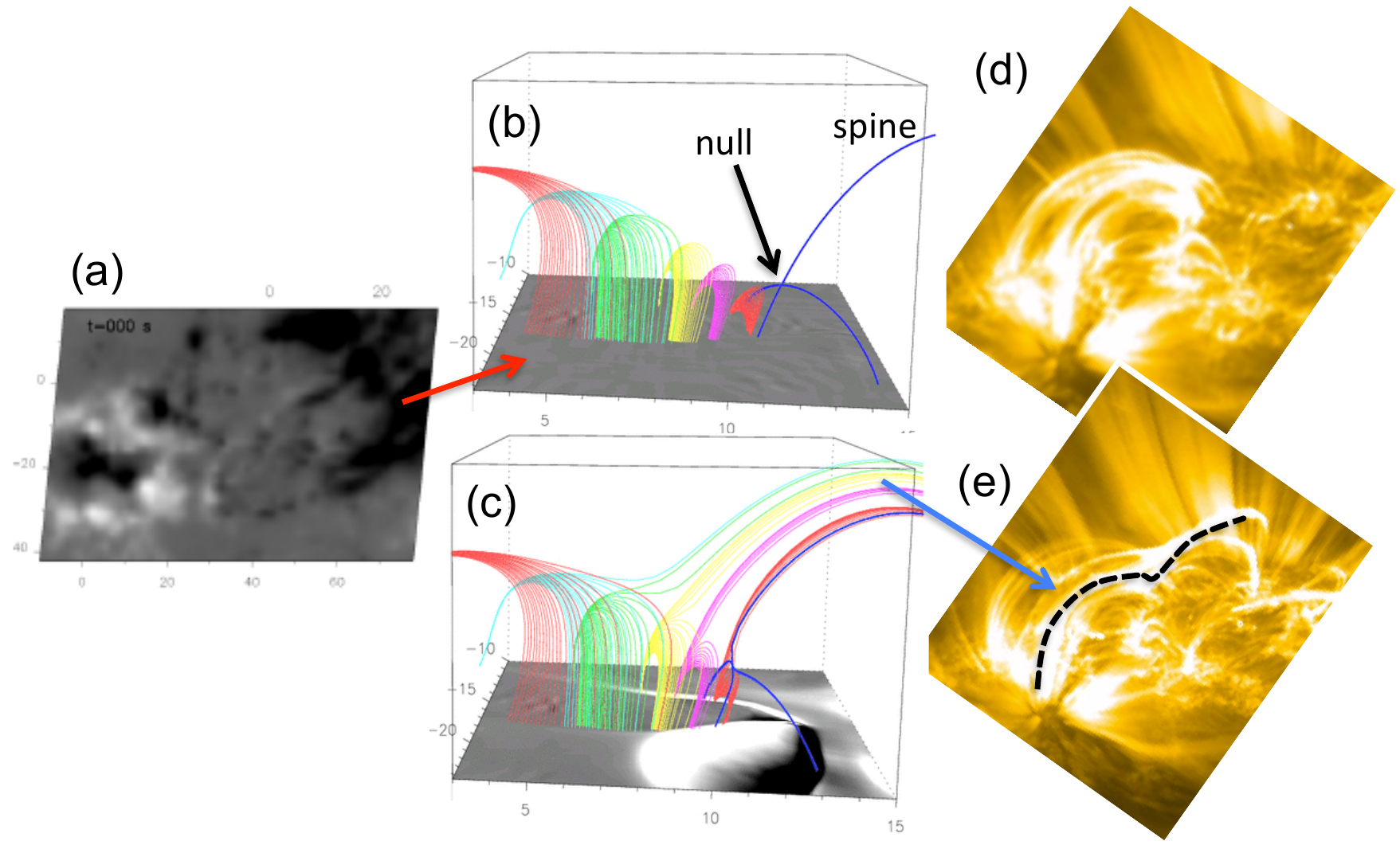} }
\caption{Example of a simulation of a confined flare. 
(a) Photospheric magnetogram on 16/11/2002 (\mm{SOL2002-11-16T13:58}) with positive/negative polarity in light/dark grey.
(b,c) Evolution of field lines in the MHD simulation before and after reconnection at the null point (OHM code, adapted from \citealt{Masson2009}). The dark-blue field lines are separatrix field lines passing through the null point. The other colored field lines are plotted with fixed footpoints in a positive flux area. The bottom boundary shows the distribution of the vertical electric current density (gray scale). 
(d,e) Comparison of the simulation to another confined flare: the flare loops before and after the occurrence of the flare appeared during the 11/03/2014 flare (\mm{SOL2014-03-11T03:50}) recorded by AIA/SDO. }
 \label{fig:confinedflare}
\end{figure}  

On the other hand, large scale flux emergence can also occur with the \mm{appearance} of new polarities emerging in older regions, without being necessarily associated with an already formed flux rope. Such conditions then lead to \mm{the appearance of} several magnetic field connectivity domains. For example, several examples \citep[\eg\ ][]{Mandrini1993, Gaizauskas1998} showed that almost oppositely oriented bipole \mm{that} emerged between the two main polarities of an AR created topological conditions for current layers to form, and as such for flaring activity to occur.
In particular, in the active region 2372 studied by \citet{Mandrini1993}, some of the \Halpha\ flaring kernels were directly associated with flux emerging processes.
Therefore, the appearance of several domains of connectivity, as well as their associated photospheric motions, are both favorable conditions for magnetic reconnection to take place.
Since flux emergence is ubiquitous and continuously takes place in the solar corona \citep{Schrijver2009}, this mechanism \mm{has been put forward as possibly an important process occurring in} the majority of flares, and is in particular a rather straightforward process to trigger confined flares.  

A test of flux emergence as a possible mechanism for confined flares as been proposed as follows. \citet{Masson2009} studied an important emergence occurring in the middle of AR 10191 and which led to a confined flare on 16/11/2002 (\mm{SOL2002-11-16T13:58}).  The photospheric motions are first modeled to reproduce the evolution of photospheric magnetograms from MDI/SOHO \citep{Scherrer1995}, and a potential magnetic field extrapolation is used as an initial condition. The MHD evolution shows the build up of localized electric currents, then the release of magnetic energy by reconnection.
This magnetic reconnection takes place at the null point and the surrounding QSLs, and leads to the formation of newly reconnected field lines (\fig{confinedflare}b,c) in good correspondence with coronal observations \citep{Masson2009}.  Note that although the simulation was defined as a case study for AR 10191, similar configurations of reconnected field lines can be found in other confined flares and now in other data-driven simulations of null-point reconnection. \fig{confinedflare}d,e represents heated coronal loops before and after the 11/03/2014 (\mm{SOL2014-03-11T03:50}) flare event obtain with the AIA instrument onboard SDO, and is presented as an illustration.

\subsection{Eruptive flares: Unstable Flux Ropes} 
      \label{sect:unstablefluxropes}      

Since eruptive flares are defined as CME-accompanying flares, there is a necessity to understand what triggers its magnetic field configuration to be ejected in the first place.  Since this structure can remain stable from days to weeks above the solar surface (see prominences/filaments in \sect{CMEs}), a CME ejection can be seen as a sudden loss of balance between two acting forces: the magnetic tension \mm{of coronal arcades} above the flux rope that pushes it downward, and the magnetic pressure of the flux rope that leads to further expansion. Then, a search for a trigger mechanism must account for either the magnetic tension to reduce or for the magnetic pressure to increase. In the review of  \citet{Aulanier2014}, the author discussed the necessary ingredients for a twisted structure to be expelled in the outer corona, which we briefly recall in the following.

Looking at the evolution of active regions or regions associated with prominences and filaments can help understand what the ingredients to be accounted for triggering scenarios are. For example, during the decaying phase of active regions the dispersion of magnetic flux by convective motions implies a converging motion at the PIL as well as flux cancellation.
\citep[\eg\ ][]{Martin1985,Schmieder2008,Green2011}.
The coronal response to such photospheric evolution has been studied in many numerical models, such as the 3D simulations of \citet{Amari2003}, \citet{Mackay2006}, \citet{Yeates2009}, and \citeauthor{Amari2011} (2011,2014). 
While flux cancellation at the PIL 
leads to magnetic flux decrease in the photosphere, it transforms the \mm{magnetic arcades overlying the flux rope - in other words, the overlapping field lines}, to field lines wrapping around this flux rope, so further building it. 
Then, the photospheric evolution leads to a decrease in the coronal magnetic tension, a favorable condition to weaken the downward force acting on the flux rope. 

  \begin{figure}  
 \centerline{ \includegraphics[width=1\textwidth]{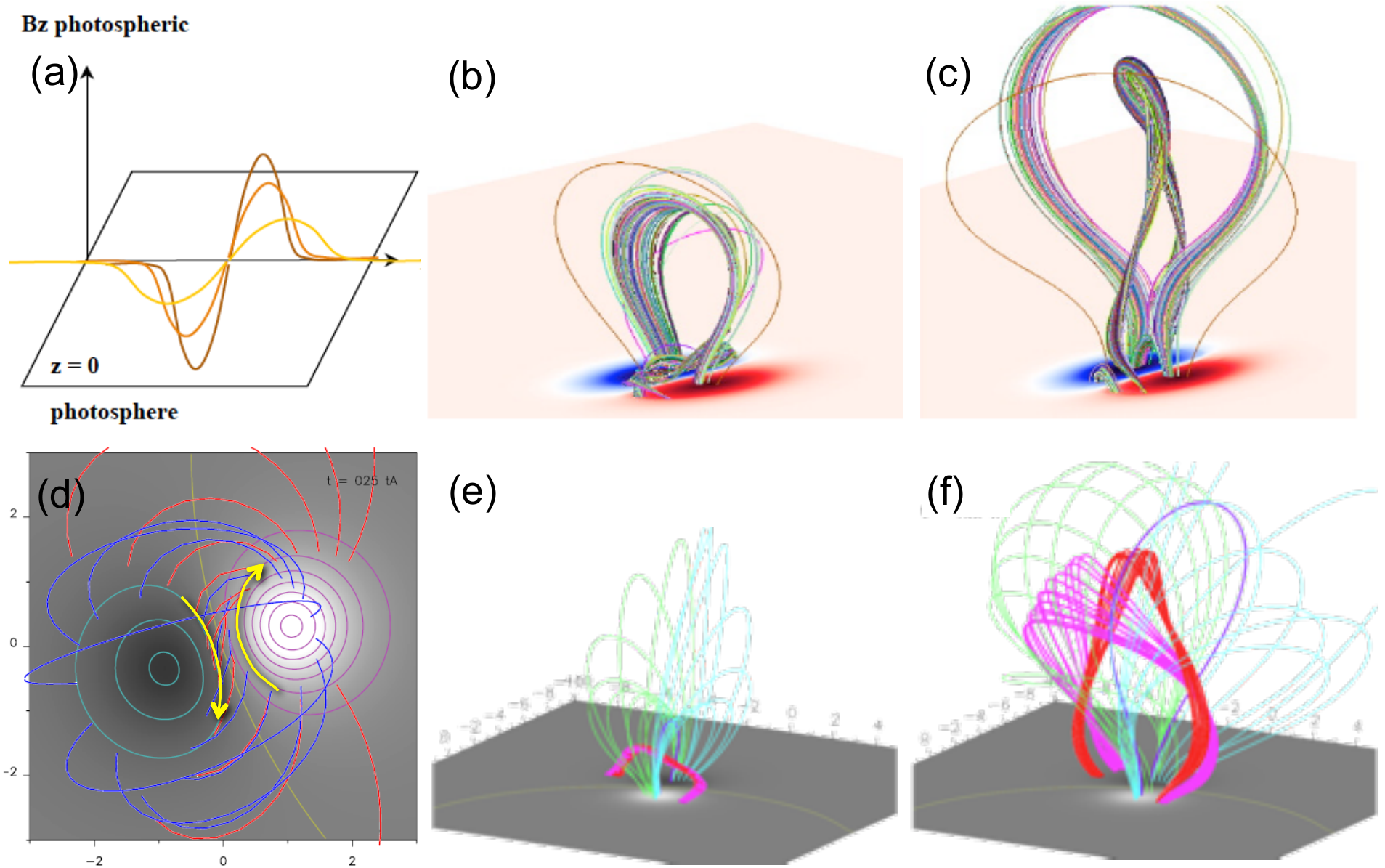} }
\caption{Examples of flux rope formation and destabilisation in 3D numerical simulations. (a-c): magnetic flux dispersal (panel a) leading to a flux rope formation then ejection (panels b and c), adapted from \citet{Amari2003b}. (d-f): flux rope formation at the bald patches following flux diffusion and shearing motion at the photosphere (yellow arrows panel d). Panels (e,f) represent two time steps before/after the flux rope has reached the torus-instability threshold, adapted from \citet{Aulanier2010}.} 
 \label{fig:FRformation}
\end{figure}  

Although several other processes can be \mm{identified as participating} in the triggering processes (such as small and large scale reconnection inducing tension weakening, flare reconnection from below, shearing motions building up pressure below the overlying arcades, flux cancellation decreasing the arcade flux -- see references in \citealt{Aulanier2014}),  a question remains about the actual intrinsic mechanism that triggers flares. Such a model would take into account the different phenomena observed  prior to flares, explaining the trigger of CME eruptions for a majority of the cases. 
Then, several authors have studied flux rope ejections with numerical simulations by bringing \mm{a flux rope} to an unstable state (\fig{FRformation}): \citet{Amari2000, Lin2001, Forbes2006, Fan2007, Aulanier2010, Olmedo2010, Zuccarello2012}. The characterisation of this instability was proposed by \citet{Aulanier2010}: from the formation of the flux rope (see \sect{Buildup}), they studied its stability by turning down the driving forces (\ie\ the photospheric boundary motions) at different times in the simulation. By doing so, they found that there is a threshold provided by the configuration of the flux rope, from which it becomes unstable. This instability is the torus instability (see also \citealt{KliemTorok2006}).

Another scenario has been proposed for the magnetic tension reduction by \citet{Antiochos1999} 
and \citet{DeVore2008}. 
In their so-called breakout model, a current sheet is formed in a quadrupolar configuration at the null point located above \mm{a sheared magnetic field configuration} (which could also host a flux rope, but not necessarily). ÊAt some point of the evolution a fast reconnection is triggered there. It removes rapidly part of the magnetic tension over the sheared arcade, inducing its eruption. Although this configuration can be observed in the solar atmosphere (see \eg\ \citealt{Aulanier2000}), it requires that too much flux cannot accumulate at higher altitude, otherwise the magnetic tension reduction via reconnection becomes too inefficient to account for the sudden ejection of CMEs. Since this scenario involves mostly sheared arcades rather than a flux rope structure, a main ingredient in the ejection of CMEs, we will focus in the following on simulations of torus-unstable flux ropes.

\section{Advances in flare QSL reconnection}
      \label{sect:eruptiveflares}

\subsection{QSLs in the Presence of Twisted Flux Tubes}
      \label{sect:QSLsfluxropes}      

 \begin{figure}  
 \centerline{ \includegraphics[width=1\textwidth]{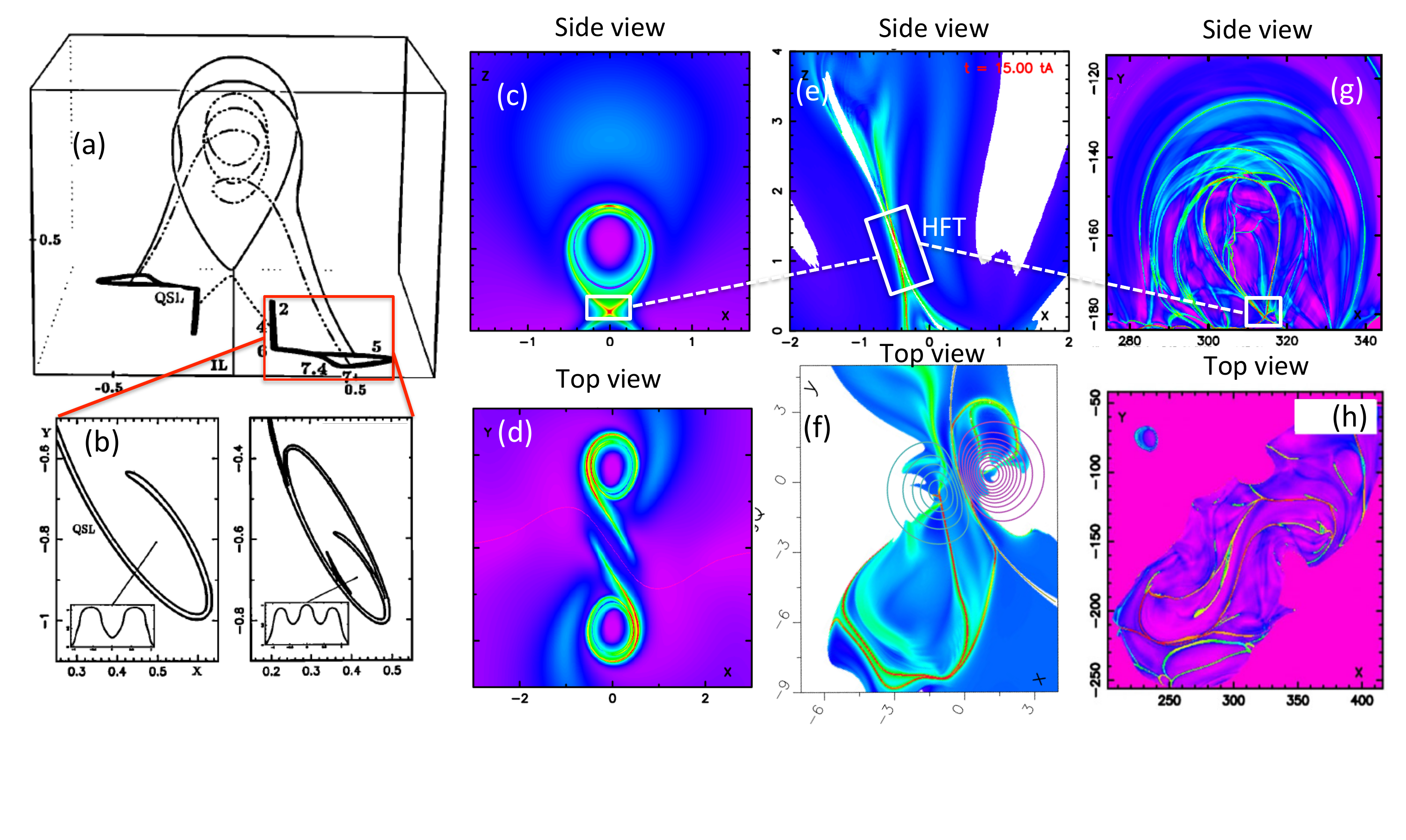} }
     \vspace{-0.08\textwidth}   
\caption{(a) QSLs footprints and field lines for an analytical model of a flux rope \citep{Demoulin1996b}, (b) with different hook shapes depending on the twist of the flux rope structure. The footpoints of the flux rope are located inside the hooks of the photospheric QSLs. (c,d) Numerical computation from \citet{Pariat2012} of the QSLs from the analytical Titov \& D\'emoulin flux rope model. The central cut (c) shows the tear-drop shape of the QSL volume surrounding the flux rope, and underneath it the HFT, where the squashing factor becomes the highest (in red). The top view (d) shows the photospheric signature with the hooks closing onto themselves.
(e) QSLs seen from a 2D cut in the 3D flux rope ejection numerical model (see \citealt{Janvier2013}) showing the tear-drop shape above the HFT, as well as the cusp structure underneath it. (f) The photospheric footprints of the QSLs show an asymmetry, due to the magnetic polarity asymmetry. (g,h) QSLs computed from the flux-rope insertion extrapolation model of the 12 February 2007 event. The QSL computation renders a much more complex structure, although the tear-drop shape, cusp and HFT are discerned in the side view, as well as the hook-shaped photospheric QSLs (from \citealt{Savcheva2012}).
} 
 \label{fig:QSLFRs}
\end{figure}  

It is necessary to understand the 3D topological features associated with flux ropes to explain the underlying processes linked to their ejection during an eruptive flare.
In the analytical work of \citet{Demoulin1996b}, it was shown that QSLs form a thin volume wrapping around flux tubes. Their mapping, down at the photospheric boundary, is representative of the twisted shape of the flux rope, with the hook part swirling more with a more twisted configuration (see \fig{QSLFRs}a,b). 
The evolution of QSLs during the formation of a flux rope by kinematic emergence through the photospheric boundary was investigated in \citet{Titov1999}. In particular, it was shown that as the flux rope fully emerges, the photospheric footprints of the QSLs split into two distinct traces, reminiscent of the \Halpha\ ribbons during flares \citep{Titov2007}. 

This analytical flux rope (see Section 2 in \citealt{Titov1999}) was reproduced numerically by \citet{Savcheva2012} for comparisons with MHD simulations and observations.
The volume associated with the highest values of the squashing degree $Q$ (see \sect{QSLs}), defined as the hyperbolic flux tube \citep[HFT,][]{Titov2002}, is located below the flux rope (see \fig{QSLFRs}c). This HFT volume maps all the way down to the photosphere to the beginning of the hook-shaped structure (see the red area in the photospheric footprints of the QSLs in \fig{QSLFRs}d). In the close vicinity of the HFT, different types of connectivities are present. Above are field lines from the flux rope, below are arcade-like field lines, and on the sides are large scale arcade-like field lines surrounding the flux rope. This variety of connectivities in a small region implies that the QSLs, and especially their core, the HFT, are privileged regions both for current layer build-up and for the occurence of magnetic reconnection.  

For configurations before eruptions, seen as on-disk sigmoid structures, it is interesting to compute the QSLs and this first requires to compute the coronal magnetic field. One of the most successful methods to reproduce the presence of a twisted coronal magnetic structure is that of \citet{VanBallegooijen2004}, who proposed inserting a flux rope in the magnetic region of a sigmoid region, extrapolated with a potential field, and leaving the system to relax using a magneto-frictional code.
This method was successfully applied to several active regions presenting sigmoids as an evidence of the presence of a flux rope \citep{Savcheva2009, Savcheva2012, Su2012}. In particular, in \cite{Savcheva2012}, the authors calculated the QSLs 
from their extrapolation method. As shown in \fig{QSLFRs}g,h, the structure of the QSLs, although more complicated due to the presence of more complex surrounding coronal loops, also show an HFT, tear-drop and cusp shapes above and beneath it, as well as hook-shaped photospheric signatures. As such, the presence of an HFT is an essential part in the sigmoidal structures seen in the corona, and as will be seen below, holds the keys to the underlying reconnection processes forming flaring structures.

From the results of the pre-eruptive magnetic field, one can predict 
that the flare reconnection will occur at the HFT. This analytical result was indeed verified in a dynamical MHD model reproducing the ejection of an unstable flux rope \citep{Janvier2013}, from the original setting of \citet{Aulanier2010} but with no photospheric boundary driving applied. 
Computing the squashing degree in the 3D volume, the QSLs are seen to wrap around the modeled flux rope (\fig{QSLFRs}e) similarly as the analytical flux rope model. As time evolves, the tear-drop shape becomes bigger as the flux rope extends, and the HFT underneath it moves upward, following the ejection motion of the flux rope, while the cusp region becomes wider. Since this cusp structure indicates the delimitation between the regions where field line have not reconnected (coronal field lines) and regions of reconnected field lines (the flux rope above the cusp, and the flare loops inside the cusp region), the cusp becomes wider as more flare loops are fed \m{into} the region. 

The photospheric signatures of the QSLs, as seen in \fig{QSLFRs}f, also present a hook-shape structure, and the straight parts near the PIL move away from each other as the eruption goes on, while the hooks become rounder. Note that the hooks do not close onto themselves, as found in the analytical Titov-D\'emoulin model shown in \fig{QSLFRs}c,d. This is because the flux rope structure is not as twisted as that in the analytical model. A comparison with the morphology of the flare ribbons (see \sect{currentevolution}) indicate that flux ropes in the corona should not have 
twist much larger than one turn, contrary to what is suggested in kink-unstable flux rope models \citep{Janvier2014}.

\subsection{Reconnection in 3D: Slipping Motion} 
      \label{sect:slipping}      

  \begin{figure}  
\centerline{ \includegraphics[width=1\textwidth]{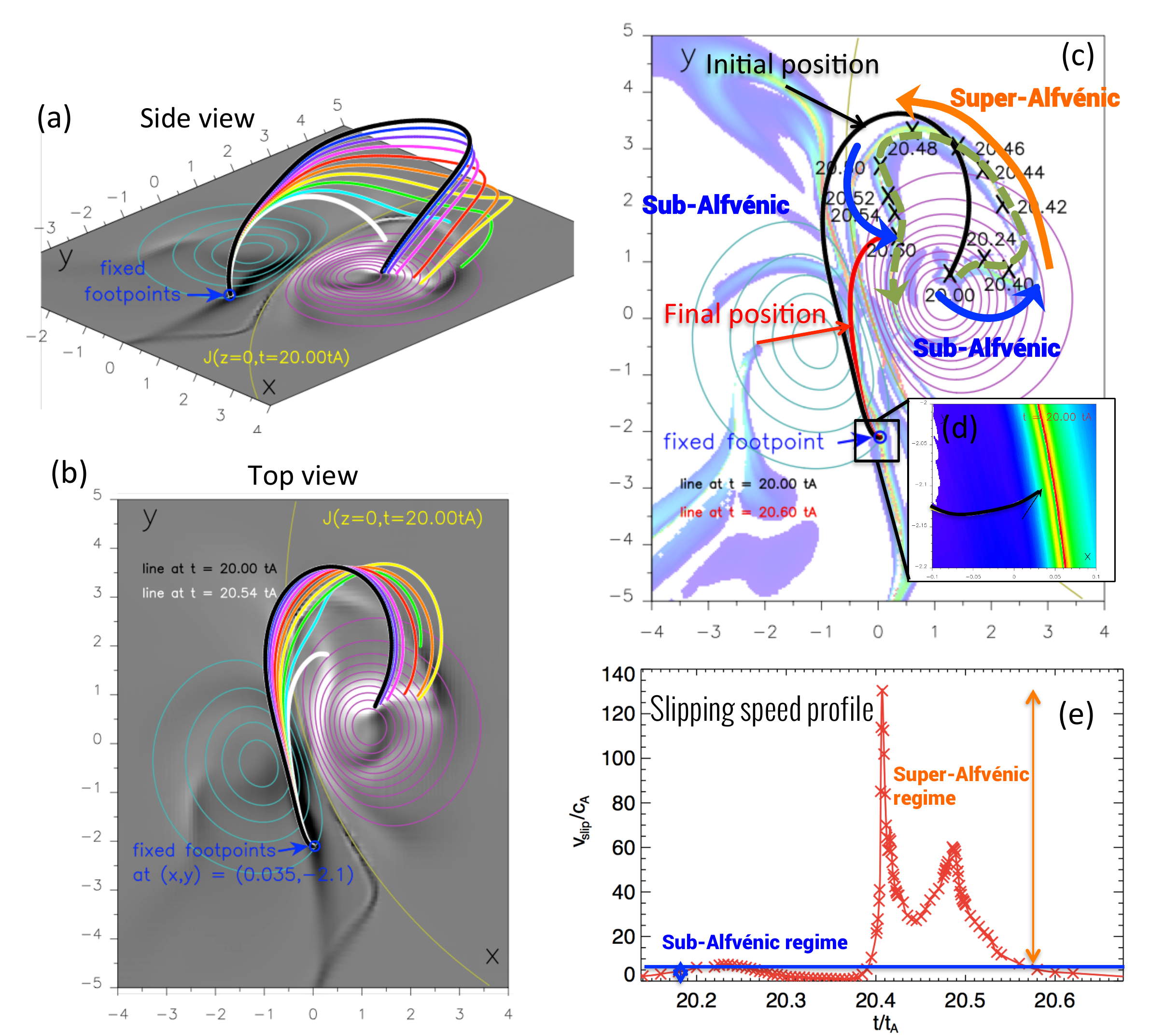} }
\caption{Deducing the speed of the slipping motion of a field line from a simulation (see \citealt{Janvier2013} for the simulation setup). (a-b): Side and top views, respectively, of the same field line plotted at different times as it reconnects with neighbouring field lines. The thick black line represents the initial position, while the change of colors from black to white represents different times in the simulation. The countourplots indicate the strength of the magnetic field for both the negative (cyan) and positive (purple) polarities, while the yellow line represents the PIL. The background plot in grayscale indicates the strength of the photospheric electric current densities, where the black (resp. white) contours indicate the strongest negative (resp. positive) $\Jz$. The positions of the moving footpoint as well as their corresponding time in the simulation are reported on a photospheric map of the QSLs (panel c), along with the initial and final position of the loop (in black and red). The regimes where the slipping speed is sub-Alfv\'enic or super-Alfv\'enic are reported as well. Sub-panel (d) shows a zoom of the photospheric footprint of the QSL surrounding the fixed footpoint region. (e) The speed of the slipping motion is reported at each timestep of the simulation, showing its non-constant evolution.  }
\label{fig:slipping}
 \end{figure}  

Quasi-separatrix layers indicate regions of strong magnetic field distortion. 
When field lines are reconnecting with their neighbouring field lines in a QSL, their motion is seen as an apparent \textit{flipping}, or \textit{slipping} motion of field lines. This was analytically predicted by several authors \citep{Priest1992, Priest1995, Demoulin1996}.

Since it is impossible to follow elementary flux tubes in the corona, numerical simulations present an invaluable resource to study the dynamics of field lines during the reconnection process. Indeed, by fixing a field line starting point, one can evaluate its successive changes of connectivity by tracing the line from its defined point, as time goes by. Note that if the change of connectivity occurs at super-Alfv\'enic time scales, the apparent slipping motion is said to be ``slip-running'' \citep{Aulanier2006}.
\m{This term refers to the fact that within the MHD paradigm information only travels up to the Alfv\'en speed in a low-$\beta$ plasma.  The} fixed footpoint of the magnetic field line \m{at one end} does not actually ``see'' the change of connectivity if its opposite footpoint moves at super-Alfv\'enic speeds \mm{this opposite footpoint is defined as the photospheric endpoint location of the same field line computed from the fixed footpoint location,} . 
At the scale of the system, for super-Alfv\'enic or slip-running motion, field lines behave as if they had reconnected in the presence of separatrices. This motion has been reported in different simulation setups \citep{Torok2009, Masson2009, Masson2012, Janvier2013}. 

\m{The} characteristics of the slipping motion of field lines during an eruptive flare were studied by \citet{Janvier2013}, and in particular the speed of the motion \m{was} quantified, for the flux rope ejection simulation shown in the bottom row of \fig{FRformation}.
An example of a slipping field line is presented in \fig{slipping} (panels a and b), where the same field line, defined by its footpoint anchored in the negative polarity, is traced at different times in the simulation. Since the \mm{location of} other footpoint is moving as the consequence of the field line reconnecting with its neighbouring field line, it is possible to report for each time step the distance travelled (\fig{slipping}c, where the photospheric QSL footprints have also been reported), and hence the slipping motion speed, that is shown in \fig{slipping}e. This slipping motion speed is not constant in time: rather, it has spikes, and its values can be sub- or super-Alfv\'enic. 

As the \mm{location of the} non-fixed footpoint moves along the QSL in the positive polarity, the other fixed point remains at the same location in the negative polarity. It is anchored outside of the QSL before the reconnection takes place, as indicated in the zoomed region in \fig{slipping}d. However, as time evolves, this QSL moves away from the PIL, and hence \mm{``sweeps''} the field line footpoint \mm{location}. Indeed, as the flux rope is ejected away, the trailing reconnecting region moves upward, so that field lines passing through it reconnect with each other and the photospheric footprints of the QSLs in each polarity are then seen to move away from each other \citep[similar to the CSHKP model,][]{Carmichael1964, Sturrock1966, Hirayama1974, Kopp1976}. As such, while the motion of the footpoint \mm{location} in the positive polarity is \textit{along} the QSL, the footpoint \mm{location} in the negative polarity has a relative motion with respect to the QSL \textit{perpendicular} to it. Then, the speed of the moving footpoint \m{location} is 
super-Alfv\'enic when the fixed footpoint is in the core of the QSL, and sub-Alfv\'enic on both sides.
Note that taking a field line that would be defined by a fixed footpoint \mm{location} in the positive polarity would give the inverse result: the fixed footpoint would move perpendicularly with respect to the QSL footprint in the positive polarity, while its corresponding displaced footpoint \mm{location} in the negative polarity would move along the QSL footprint. Then, the slipping speed of the moving footpoint \mm{location} is directly related to the distortion of the magnetic field, given by the field line mapping ($N$, see Eq.(4) in \citealt{Demoulin1996}) and the speed at which the QSL is displaced: $v_{\rm{slipping}}=N\times v_{\rm{QSL}}$ (see \citealt{Janvier2013}).

From its first prediction in the 1990s \citep{Priest1992, Priest1995, Demoulin1996} and its numerical modelling \citep{Aulanier2005,Torok2009,Masson2009,Janvier2013}, the slipping motion of field lines has been observed for the first time in coronal loops  \citep{Aulanier2007} and during eruptive flares \citep{Sun2013,Dudik2014,Li2014} very recently. These observations, at high temporal cadence, can only provide evidence of sub-Alfv\'enic motions, since it is the only regime for which the response of the plasma to the change of connectivity can be observed.
The slipping field lines of the coronal loops \citep{Aulanier2007} was monitored with the XRT instrument onboard HINODE, and were seen to exchange their footpoint positions with a motion speed estimated between the 30 to 150 km/s.
\citet{Dudik2014} reported the slipping motion of field lines in the X-class eruptive 1 July 2012 (\mm{SOL2012-07-12T16:50}) with the AIA/SDO instrument, and also showed the propagation of kernel brightenings along the flare ribbons, reminiscent of the motion of footpoints obtained in the MHD simulation of a similar-looking active region \citep{Janvier2013}. The authors reported on apparent velocities ranging from 4 km/s to 140 km/s, and the successive kernel brightenings were associated with the footpoints of the field lines forming the envelope of the flux rope that eventually gets ejected during the flare (a CME was indeed recorded for this event). 
 Note that since this process should be in principle intrinsic to any flare (as they all have a 3D structure), it is not related to the particular zipping brightening motion reported in some flares \citep{Tripathi2006}. The latter can be explained by the propagation of the reconnection site in one direction due to an asymmetric filament eruption, in which case both slipping motion and zipping motion can take place, although not necessarily on the same timescales.

\subsection{Electric Current Evolution} 
      \label{sect:currentevolution}      

\begin{figure}  
\centerline{ \includegraphics[width=1\textwidth]{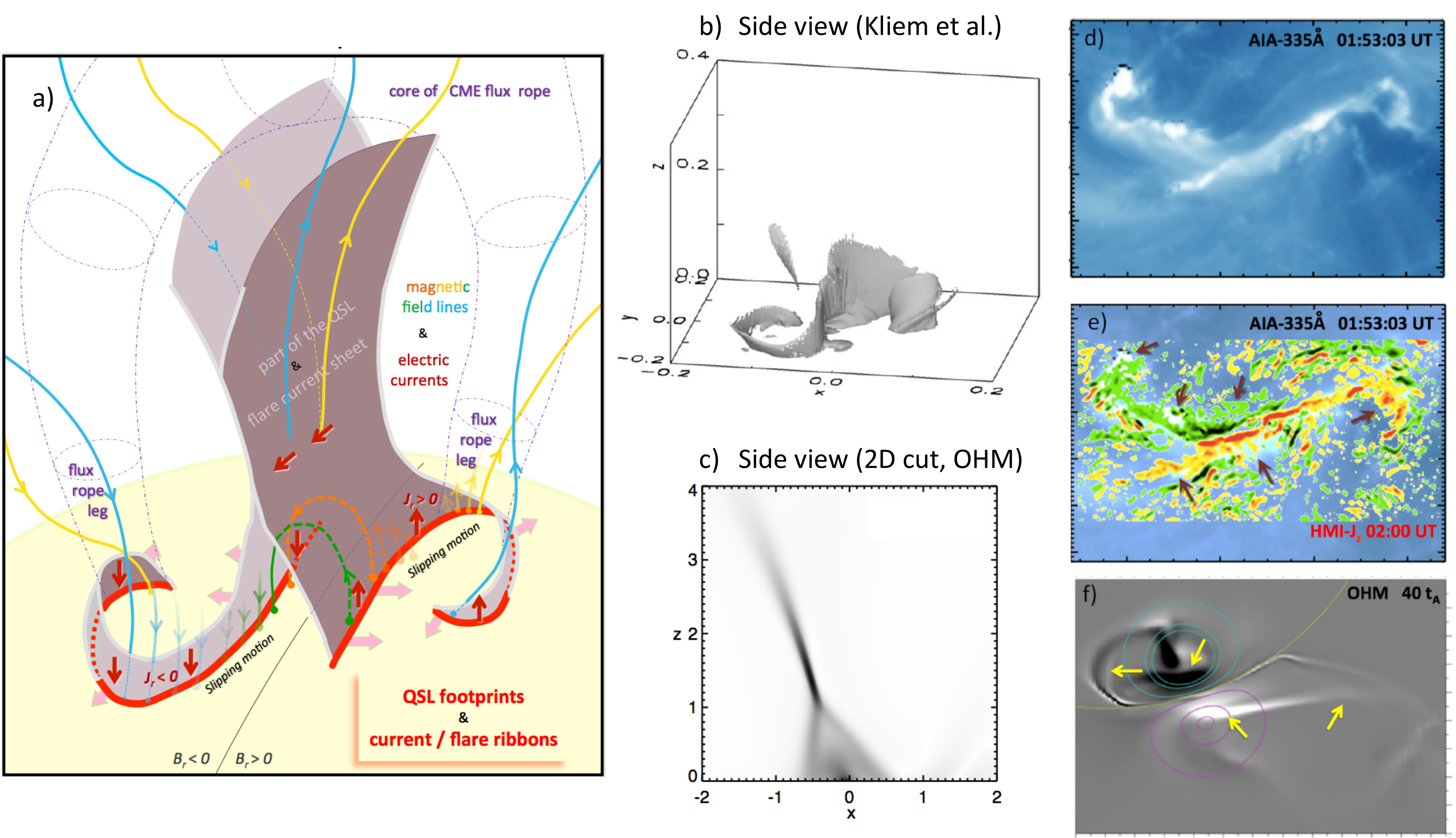} }
\caption{(a) The 3D standard model for eruptive flares \citep{Janvier2014} derived from the interpretation of eruptive flares with simulations of \m{a} torus-unstable flux rope. The grey surface indicates the QSL volume that wraps around the flux rope structure (purple dashed lines), as well as the current density volume that maps \m{to a} similar location (although it does not extend above the flux rope, see panel c). The coloured field lines represent coronal loops that have slip-reconnected within the QSL, as indicated by the series of footpoints representing the slipping motion. The blue and yellow loops form the outer envelope of the flux rope, while the green and orange loops represent the newly formed flare loops. The footpoints of the flux rope are located within the hook of the QSL/current volume photospheric footprints, while the footpoints of the flare loops are located in the region delimited by the straight portion of these footprints. (b) Isosurface of (10\%) the maximum value of the current density, in the simulation of data-driven unstable flux rope \citep{Kliem2013}, showing its extension within the simulated volume and its mapping all the way down to the photospheric boundary, and showing the hook structure wrapping around the flux rope. (c) 2D cross-section of the current density volume, taken at the central part of the flux rope, in a torus-unstable flux rope ejection model \citep{Aulanier2012}, showing the thin intense current portion (at a similar, though not exact location as the HFT, see \citealt{Janvier2013}), as well as the cusp shape similar to that found for QSLs (see \fig{QSLFRs}). (d) Flare ribbons seen in the 335~\AA\ filter of the AIA/SDO instrument during the 15 February 2011 X2.2 flare (\mm{SOL2011-02-15T01:56}), showing a similar $J$-shaped structure as the photospheric current ribbons and QSLs. The intense emission locations have been compared with the location of strong electric currents derived from the HMI magnetograms, as indicated with the brown arrows in panel e. (f) Current ribbons in the OHM simulation of an eruptive flux rope, showing similar changes (intense patches of current density, outward motion and hooks becoming rounder) as those found for current and flare ribbons from the observations (panels d and e). }
\label{fig:currentsmodels}
 \end{figure}

The formation of electric current layers is an essential component of flaring activity. These thin layers are predicted to form in the regions where the magnetic field is strongly distorted, \ie\ at similar locations as the Quasi-Separatrix Layers (see \sect{QSLs}). On the other hand, as shown in \fig{QSLFRs}, QSLs form a coronal volume that maps all the way down to the photospheric boundary, with a photospheric signature presenting a $J$-shape. Such a structure can be described in a 3D model of eruptive flares, where the core of the CME (\ie, the flux rope) fills a coronal volume embedded in the QSLs (see \fig{QSLFRs}) mapping in the hooks of the QSL footprints, flare loops  are embedded in the cusp region, and field lines are slipping within the QSL. The representation of such a model is given in \fig{currentsmodels}, panel a, where these main characteristics have been put together. Then, do current layers have a similar morphology as that of the QSLs?

3D numerical modelling of eruptive flares has been able to answer this question by mapping the regions of high current densities. For example, in \citet{Kliem2013}, the authors used an extrapolated field of the AR 11060 as an initial condition to run their numerical code. 
The associated 3D current layer for the present model\mm{, obtained as a current density itself derived from the components of magnetic field obtained in the simulation}, is shown in \fig{currentsmodels}b. \mm{It }presents a similar morphology as for the QSLs, although the high current density layer does not wrap around the flux rope as the QSLs do. This 3D view also shows the rounded hooks that \m{surround} the core region of the flux rope. In \citet{Janvier2013}, the authors studied the evolution of the current layer as time goes by: a 2D cut in the central part of the flux rope 3D volume clearly shows the cusp region as well as the current layer above it where reconnection takes place. Note however that the region of highest density, found at the cusp in the 2D cut, is slightly offset compared with the location of the highest squashing factor location, namely the HFT (see Figure~2 of \citealt{Janvier2013} for comparisons). 
As the flux rope expands, this current layer moves upward and becomes thinner, while the hooks of the photospheric current layers (shown for example in \fig{currentsmodels}f) become rounder and the straight parts of each current ``ribbon'' separate from each other, away from the PIL, as more flux is fed into the flux rope structure. As such, the 3D view contains the evolution of the field as represented by the 2D model. These modelled photospheric changes in electric current resemble those of observed bright flare ribbons during eruptive flares (\eg\ \fig{currentsmodels}d). \mm{As such, the model for eruptive flares shown in \fig{currentsmodels}a incorporates the different evolution ingredients above. It proposes a development from the 2D point of view of the standard model of eruptive flares in 2D (or CSHKP model) to its more complete and more accurate 3D version \citep[see][]{Janvier2014}.}

\mm{A question that remains to be answered from the description above is how the flare ribbons and simulated photospheric current ribbons are related.} 
Using the measurements of the three components of the magnetic field at the solar surface means that it is possible to observe the evolution of the photospheric current density, \m{provided that there is} a good spatial and temporal resolution. The HMI instrument onboard the Solar Dynamics Observatory has an unprecedented temporal resolution, providing maps of the line-of-sight measurements of the magnetic field every 12 mn. 
In \cite{Janvier2014}, the authors used the Milne-Eddington inversion code UNNOFIT \citep[see][]{Bommier2007} applied to the HMI magnetograms covering the active region NOAAA 11158 during 4 hours of observation, including the X2.2 flare that took place on 15 February 2011 (\mm{SOL2011-02-15T01:56}) and peaking at 01:56 UT.
From this inversion process, maps of the vertical component of the current density vector $\Jz$ were deduced. Note that although the active region presents a quadrupolar configuration, the X-class flare of Feb. 2011 was mainly confined \m{within} the two central bipoles, and as such had a configuration (including the asymmetry in the strength of the magnetic polarities) similar to that studied previously in the eruptive flare model of \citet{Aulanier2012}.

\m{Investigating the evolution of $\Jz$} at different times before and after the peak of the flare, the authors reported an increase in the direct electric current component along both the current ribbons in the negative and positive polarities (in regions indicated with arrows in \fig{currentsmodels}e). These current ribbons delimitate the region around the PIL where the magnetic field becomes more and more horizontal, showing that these regions do correspond to the formation of new post-flare loops \citep{Sun2012, Wang2012,Petrie2013}.

These results seem \textit{a priori} contradictory to the core idea that flares typically dissipate the stored free energy as the magnetic configuration evolves toward a more potential, and hence less current-carrying, one. While a strong dissipation of magnetic energy is expected in the thin current layers, at the same time the electric current density increases due to the collapse of the thin layers. 
Such an evolution is due to the upward \m{ejective} motion of the torus-unstable magnetic flux rope: The field lines underneath the flux rope get squashed in the trailing current region (\fig{currentsmodels}c), which in turn thins while the current density increases there. Since this current volume maps all the way down to the photosphere (panel b), changes are then also expected to be seen at the photospheric level, as predicted in the numerical simulation (panel f). \m{So} although this mechanism competes with the strong diffusion taking place, the total current density is large at the beginning, during the flare impulsive phase, before decreasing later on. 

The location and the evolution of enhanced electric currents within flare ribbons, predicted with numerical simulations, have now been confirmed thanks to the enhanced temporal and spatial resolution of recent instruments. They are not only correlated to the evolution of flare ribbons, as summarised above, but also to a variety of other phenomena: they map at similar locations as strong twist of the magnetic field, as shown in a data-driven simulation of the same X2.2 flare of 15 February 2011 (\mm{SOL2011-02-15T01:56}) by \citet{Inoue2014}, and regions of current changes have also been associated with HXR sources observed with RHESSI \citep{Musset2015} as well as sunquakes \citep{Zharkov2011}.

\section{Conclusion} 
      \label{sect:Conclusion} 
 
The present paper summarizes the MHD building blocks of solar flare modelling, from the topology to the trigger mechanisms, and from observations to 3D numerical simulations.
While a flare is typically classified by its strength depending on its soft X-ray emission, it can also be differentiated in two categories, either confined or eruptive (see \sect{Observations}). While confined flares do not have a large influence on the global corona, eruptive flares are typically associated with coronal mass ejections (CMEs, see \sect{CMEs}), and can therefore be potentially disruptive for surrounding planetary environments. 
They share similar features, such as the existence of ribbons and flare loops (\sect{flareloops}).

Observations of solar flares inevitably lead to the question of the storage and the release of the magnetic energy that fuels them. Several mechanisms have been proposed (\sect{Energy}): one scenario involves current-carrying magnetic structures that form underneath the photospheric boundary, in the convection zone, and that are transported all the way to the corona. 3D simulations that have been developed recently are able to reproduce this emergence process, sometimes all the way to the ejection of current-carrying structures such as flux ropes (see \sect{Buildup}). Other scenarios involve the formation of non-potential fields by photospheric motions such as shearing or flux cancellation, and have been tested against observations of the evolution of active regions and coronal loops.

Magnetic reconnection has been proposed as a mechanism to dissipate the free energy (\sect{reconnection}). This phenomenon involves the creation of thin layers of high current density where the plasma becomes non-ideal, and where the magnetic energy can be converted into heat and kinetic energy, whilst forming a population of non-thermal particles. Although the spatial scales at which reconnection is expected to occur are too small for the present-day instruments \m{to detect}, its consequences have been witnessed in numerous cases, with an ever-increasing resolution (both temporal and spatial, \fig{recoevidences}) over the years, and \m{support} the present theory.
The regions of non-idealness can be found by searching for the topological characteristics of the magnetic field, such as the presence of null points, separatrices, separators, and also Quasi-Separatrix Layers (see \sect{Toporeco}). 
Current density layers, where the magnetic energy is dissipated, preferentially form at QSLs. Searching them in coronal magnetic fields has become a powerful and successful tool to understand the signatures of flare evolutions.

Using observations as well as numerical simulations, several major mechanisms can \m{account} for confined and eruptive flares. 
Emerging flux (\sect{emergingflux}) \mm{may possibly account} for a great number of confined flares\mm{, although further understanding of the process driving them is required}. Data-driven MHD numerical simulations can indeed explain such flares. Other observations of active regions have shown that several features appear prior to the onset of an eruptive flare, such as flux cancellation and flux dispersal, as well as shearing motions of the polarities. These mechanisms can provide the conditions for a flux rope to be formed. As discussed in the review by \cite{Aulanier2014}, those conditions can ultimately lead to a flux rope ejection via the torus-unstable flux rope model (\sect{unstablefluxropes}). Next, we saw how the QSLs fill a thin coronal volume that maps all the way to the photosphere in the presence of a flux rope (\sect{QSLsfluxropes}). 
In their presence, the magnetic field lines change their connectivity successively, they are said to be slipping as time goes by (\sect{slipping}). This motion, first proposed as an analytical by-product of the presence of QSLs, has been investigated in detail in the numerical simulation of a flux rope ejection \citep{Janvier2013}.  The slipping motion of field lines is intrinsic to the reconnection in 3D, and therefore, to all flares. Apparent motions of field lines and their associated kernels indeed have been observed with high-temporal resolution imaging.

Magnetic reconnection dissipates the magnetic energy in highly localised regions of high-current density, at the QSLs. Using numerical simulations of flux rope ejection, different study show a similar behavior and geometry: the coronal current layer where reconnection takes place form a volume that maps to the photosphere, 
following \m{the QSL mapping}. 
The impulsive phase of an eruptive flare is then dictated by the sudden thinning of the layer, where the electric current density increases, allowing reconnection to take place on a short time scale. This change of regimes has been confirmed by direct measurements of the photospheric vertical currents, with a high \m{temporal-cadence instrument such as the} HMI onboard SDO. This underlines the importance of using numerical models as they can also pinpoint consequences of such mechanisms that remain difficult to observe. They therefore provide the physical understanding to interpret observations, as well as the ideas for where and how to look at them. Taking into account the latest developments of numerical flare modelling, we report in \fig{currentsmodels}, panel a, on the extension of a 3D flare model from the previous CSHKP cartoon that remains valid only in 2D.

As discussed in the paper, MHD simulations of flares have been able to reproduce observational features, help to interpret the underlying mechanisms, and make predictions. Nonetheless, several issues still remain to be addressed, with further developments needed in the future. Indeed, within the MHD framework, the kinetic physics of the magnetic energy dissipation within the 3D current layer cannot be addressed. The length scales that account for this dissipation (the ion or electron length scales) are several orders of magnitude smaller than the typical lengthscale of coronal loops. As such, approximating the physics that takes place within the current layer (such as turbulence processes, plasmoid formation, Hall effect, ion heating, kinetic effects) is one solution, given the computer power available as of today, along with the development of hybrid codes. This also means that the MHD framework needs to be extended, so as to explain how a population of non-thermal particles is created. Recent developments in laboratory plasma physics, and in particular of the partition of energy between the different species \citep{Yamada2014} may shed light on the processes happening in the corona. Furthermore, knowing the energy partition given to non-thermal particles is of high importance to understand how the energy is deposited from the corona to the photosphere \citep{Milligan2014}. 

In conclusion, the correspondences between the 3D MHD models developed in the recent years, and the observations obtained with an increasing time and spatial resolution, provide strong evidence that several physical ingredients important to flares are included. 
Such models are of great importance not only to understand the flaring activity of our hosting star, but also to interpret data of other flaring stars, for which observations remain limited.

\begin{acks}
M.J. acknowledges partial funding from the SOC of the ``Solar and stellar flares: observations, simulations and synergies" conference, held from 23-27 June 2014, where this work was presented. The authors would like to thank the anonymous referee and Lyndsay Fletcher for their useful comments that helped improve the article.
\end{acks}

   
\bibliographystyle{spr-mp-sola}

\bibliography{Flares-Prague}  


\end{article} 

\end{document}